\documentclass[twocolumn, twocolappendix]{aastex631}

\shorttitle{Astronomaly Protege}
\shortauthors{Lochner \& Rudnick}

\usepackage{multirow}

\newcommand{\astronomaly}{\textsc{astronomaly}}
\newcommand{\protegefull}{\textsc{astronomaly:\! protege}}
\newcommand{\codelink}{https://github.com/MichelleLochner/astronomaly}
\newcommand{\fileslink}{https://github.com/MichelleLochner/mgcls.protege}
\newcommand{\webpagelink}{https://michellelochner.github.io/mgcls.protege}
\newcommand{\protege}{\textsc{protege}}
\newcommand{\pybdsf}{\textsc{pybdsf}}
\newcommand{\opencv}{\textsc{opencv}}
\newcommand{\astropy}{\textsc{astropy}}
\newcommand{\muJy}{$\mu$Jy beam$^{-1}$}

\newcommand{\recall}{recall$_{100}$}
\newcommand{\fullcatsize}{6161}
\newcommand{\evaluationcatsize}{1031}
\newcommand{\randomsubset}{3500}
\newcommand{\originalcatsize}{608595}

\newcommand*\rot{\rotatebox[origin=c]{90}}
\newcolumntype{P}[1]{>{\centering\arraybackslash}p{#1}}

\newcommand{\sauron}{37}
\newcommand{\sauronlike}{146}
\newcommand{\tail}{3}
\newcommand{\xclean}{237}
\newcommand{\xdouble}{255}
\newcommand{\xlopside}{110}

\begin{document}

\title{Astronomaly Protege: Discovery Through Human-Machine Collaboration}
\correspondingauthor{Michelle Lochner}
\email{dr.michelle.lochner@gmail.com}

\author[0000-0003-2221-8281]{Michelle Lochner}

\affiliation{Department of Physics and Astronomy, University of the Western Cape, Bellville, Cape Town, 7535, South Africa}
\affiliation{South African Radio Astronomy Observatory, Liesbeek House, River Park, Liesbeek Parkway, Mowbray 7705, South Africa}

\author[0000-0001-5636-7213]{Lawrence Rudnick}
\affiliation{Minnesota Institute for Astrophysics, University of Minnesota, 116 Church St SE, Minneapolis, MN 55455, USA}

\begin{abstract}
Modern telescopes generate catalogs of millions of objects with the potential for new scientific discoveries, but this is beyond what can be examined visually. Here we introduce \protegefull{}, an extension of the general purpose machine learning-based active anomaly detection framework \astronomaly{}. \protege{} is designed to provide well-selected recommendations for visual inspection, based on a small amount of optimized human labeling. The resulting sample contains rare or unusual sources which are simultaneously as diverse as the human trainer chooses and of scientific interest to them. We train \protege{} on images from the MeerKAT Galaxy Cluster Legacy Survey, leveraging the self-supervised deep learning algorithm Bootstrap Your Own Latent to find a low-dimensional representation of the radio galaxy cutouts. By operating in this feature space, \protege{} is able to recommend interesting sources with completely different morphologies in image space to those it has been trained on. This provides important advantages over similarity searches, which can only find more examples of known sources, or blind anomaly detection, which selects unusual but not necessarily scientifically interesting sources. Using an evaluation subset, we show that, with minimal training, \protege{} provides excellent recommendations and find that it is even able to recommend sources that the authors missed. We briefly highlight some of \protege{}'s top recommendations, which include X- and circular-shaped sources, filamentary structures, and one-sided structures. These results illustrate the power of an optimized human-machine collaboration, such as \protege{}, to make unexpected discoveries in samples beyond human-accessible scales.

\end{abstract}



\section{Introduction}
\label{sec:intro}
Machine learning (ML) has become a critical tool in modern astronomy as datasets from new telescopes and surveys increase in size and complexity. In particular, it has the potential to enable the discovery of rare or new classes of sources, finding these ``needles'' in enormous ``haystacks''. \citet{Bohm2023} used ML to discover unusual sources such as supernovae and LINERS among thousands of spectra from SDSS, while \citet{Webb2020} used clustering and anomaly detection to discover new variables and flare stars in tens of thousands of light curves. However, for these algorithms to be effective the dimensionality of the input data must be reduced to a smaller set of meaningful numbers. This feature extraction is typically the most challenging step for most machine learning applications.

Deep learning \citep{Zhao2024} has revolutionized machine learning because, given enough data, it can perform highly effective feature extraction automatically, extracting a useful lower dimensional representation of complicated high-dimensional data such as images. In radio astronomy, \citet{Mostert2021,Gupta2022} and \citet{Vantyghem2024} used self-organizing maps, an algorithm based on neural networks, to discover rare sources in LOFAR \citep{LOFAR}, ASKAP \citep{ASKAP}, and VLASS \citep{VLASS} data respectively. \citet{Ralph2019} used an autoencoder to find a representation of the radio images on which to perform unsupervised clustering.

Several recent works \citep{Walmsley2022b, Slijepcevic2023, Mohale2024, Andrianomena2024, Riggi2024} have introduced the use of modern self-supervised learning techniques to automatically learn general representations for later downstream tasks, including anomaly detection.

 However, anomaly detection alone is not sufficient in the face of large datasets because many anomalies are not scientifically interesting and what is considered ``interesting'' varies between scientists. \citet{Lochner2021} introduced \astronomaly{}, a general purpose anomaly detection framework which uses active learning, a method of obtaining a small number of strategically chosen labels from a human user, to refine the anomaly detection algorithm to focus on more interesting sources. \astronomaly{} was used to discover an unusual radio galaxy in MeerKAT \citep{MeerKAT} data, called a Steep and Uneven Ring of Non-thermal Radiation \citep[SAURON,][]{Lochner2023}. An extension of the ideas of \astronomaly{} was explored in \citet{Sadr2022} where they iteratively retrained a convolutional neural network with active learning, effectively turning an anomaly detection problem into a classification problem.

 Our goal in this work is to use \astronomaly{} to find sources that we find scientifically interesting in the MeerKAT Galaxy Cluster Legacy Survey \citep[MGCLS]{Knowles2022}. To achieve this goal, we first extract meaningful features from the data using the self-supervised deep learning algorithm Bootstrap Your Own Latent \citep[BYOL,][]{Grill2020}, designed to automatically learn good representations of images. 
 
We then use a second phase of machine learning to identify, from these features, a sample of diverse, rare, and ``scientifically interesting'' sources for further follow-up. We introduce here an extension to \astronomaly{} called \protege{}, based on the work of \citet{Walmsley2022}, which is not a blind anomaly detection algorithm. Rather, \protege{} uses active learning to recommend sources of interest to a human trainer, based on a minimal, optimal set of user-provided scores. We will show that, although \protege{} can be used to find very specific source classes, by training it on a broad range of sources exhibiting unusual morphology, it is capable of learning a somewhat more general concept of what is interesting.

We begin this paper with an introduction to the broad landscape of machine learning in astronomy in \autoref{sec:ml_intro}, allowing us to place our work in context for the overall astronomy community. \autoref{sec:data} then provides an overview of the MGCLS data and preprocessing techniques. \autoref{sec:features} discusses the feature extraction method we use. We briefly explore in \autoref{sec:interestingness} practical aspects of what is meant by ``scientifically interesting''.
In \autoref{sec:anomaly}, we introduce \protegefull{} and evaluate its effectiveness with a small hand-scored subset of data in \autoref{sec:evaluation}. 

Finally, we apply \protege{} to the full MGCLS dataset and examine the types of recommendations \protege{} makes in \autoref{sec:full_dataset}, with a discussion and conclusions in Sections \ref{sec:discussion} and \ref{sec:conclusions}. The Appendix contains a mosaic of random sources to provide a comparison with \protege{}'s performance, more detailed technical information on the hyperparameter optimization, and a numerical evaluation of \protege{}'s performance and comparison with other techniques.

We include the BYOL code for extracting features and the \protege{} algorithm in a new release of the \astronomaly{} package\footnote{\url{\codelink}} and release the catalog of \protege{}'s recommendations, with strong caveats for its use.\footnote{\url{\fileslink}} Images of all the sources in the catalog, ordered by \protege{} score, can also be found online.\footnote{\url{\webpagelink}}

% \begin{nolinenumbers}
\begin{table*}
\centering
\begin{tabular}{|p{0.05\columnwidth}|p{0.5\columnwidth}|p{0.5\columnwidth}|p{0.5\columnwidth}|p{0.5\columnwidth}|}
    \hline
    & \raisebox{-0.5\totalheight}{Problem type} & \raisebox{-0.5\totalheight}{Illustrative UMAP} & \raisebox{-0.5\totalheight}{Problem type} & \raisebox{-0.5\totalheight}{Illustrative UMAP}  \\
    \hline
    
    \rot{\hspace{-4cm} Many labels} & 
    \vspace{0.1cm}
    {\bfseries Classification} - Supervised learning problem where a substantial number of examples of objects of known classes are available and an algorithm is trained to classify new examples. & 
    \vspace{-0.1cm}
    \includegraphics[height=4.5cm]{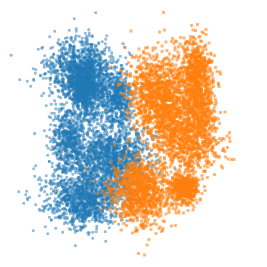} &
    \vspace{0.1cm}
    {\bfseries Regression} - Supervised learning problem where a substantial number of objects with a known corresponding quantity are available and an algorithm is trained to predict this quantity as a function of features. & 
    \vspace{-0.1cm}
    \includegraphics[height=4cm]{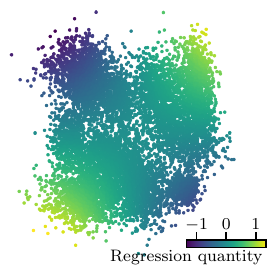} \\
    
    \hline
    \rot{\hspace{-4.3cm} No labels} & 
    \vspace{0.1cm}
    {\bfseries Clustering} - The unsupervised learning equivalent of classification where all underlying classes are unknown and an algorithm attempts to identify classes by clustering similar objects together. Researchers must label classes after they have been identified. & 
    \vspace{-0.1cm}
    \includegraphics[height=4.5cm]{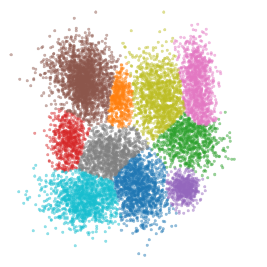} &
    \vspace{0.1cm}
    {\bfseries Anomaly detection} - Unsupervised learning problem which aims to identify anomalies - objects which appear substantially different from the norm in feature space. Many of these will not be scientifically interesting.  & 
    \vspace{-0.1cm}
    \includegraphics[height=4.5cm]{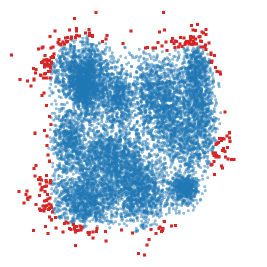} \\
    
    \hline
    \multirow{2}{*}{\rot{\hspace{-9cm} Few labels}} & 
    \vspace{0.1cm}
    {\bfseries Similarity search} - Technique to rapidly identify similar objects to a known exemplar by locating nearby objects in feature space. & 
    \vspace{-0.1cm}
    \includegraphics[height=4.5cm]{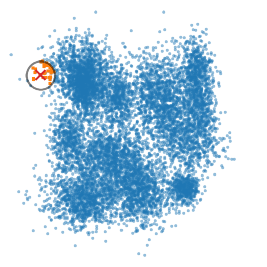} &
    \vspace{0.1cm}
    {\bfseries Classification with active learning} - A small number of difficult-to-classify objects are strategically selected for human labeling in order to improve the classifier. & 
    \vspace{-0.1cm}
    \includegraphics[height=4.5cm]{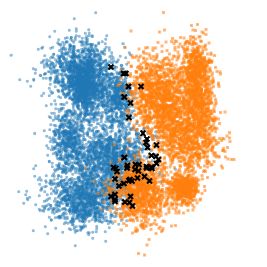} \\
    
    \cline{2-5}
     & 
    \vspace{0.1cm}
    {\bfseries Active anomaly detection} - Uses a small number of human-supplied labels (black crosses) to improve anomaly detection by excluding regions of feature space considered by the user to be uninteresting (gray points), leaving only the most interesting anomalies (red squares). This is the basis for \astronomaly{}. &
    \vspace{-0.1cm}
    \includegraphics[height=4.5cm]{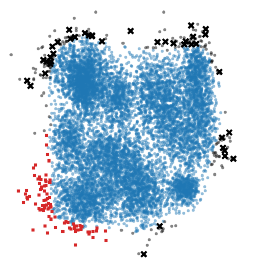} & 
    \vspace{0.1cm}
    {\bfseries Astronomaly: Protege} - Uses active learning to recommend sources which may be of interest to the user through iterative labeling and retraining of a regression algorithm, without running any initial anomaly detection algorithms. & 
    \vspace{-0.1cm}
    \includegraphics[height=4cm]{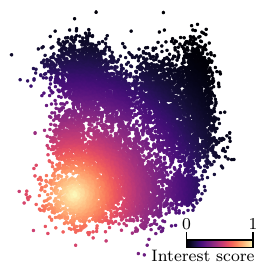}  \\
    \hline

\end{tabular}
\caption{A summary of different problems in astronomy which can be solved with different machine learning techniques. Each example is illustrated with an example UMAP plot (see \autoref{sec:visualization}) which is a low dimensional embedding of a high dimensional data space where each point represents a different image or sample in the dataset.}
\label{tab:awesome_table}
\end{table*}
% \end{nolinenumbers}

\section{PROTEGE's place in Machine Learning}
\label{sec:ml_intro}
A dizzying array of machine learning algorithms have been applied to a wide variety of problems in astronomy. Both the choice of specific technique and the amount of human labeling required will depend on the scientific problem to be solved. We first discuss the common, critical step of feature extraction, and then present \autoref{tab:awesome_table} to help understand  \protege{}'s unique role in this universe of algorithms. 

\subsection{Feature extraction}
\label{sec:feature_extraction_intro}
Machine learning algorithms in general cannot work directly with complex data such as images, light curves or spectra. These must be simplified into a lower dimensional, but still informative, representation of the data in a process called feature extraction. The choice of feature extraction method is crucial as it dictates the success of the downstream machine learning algorithm. The popularity of deep learning algorithms \citep[e.g.][]{LeCun2015}, which rely on many-layered neural networks, is due to their ability to learn useful features automatically. \emph{Self-supervised learning} is a branch of machine learning that automatically learns a representation that maintains similarity between sources in the lower dimensional space, without needing any labels for the data. It is self-supervised learning that has unlocked the potential of anomaly detection in complex image data but also necessitated the development of a new approach to anomaly detection, namely \protege{}, discussed below.

The behaviors of the different machine learning algorithms can be captured succinctly in this feature space.  To do this, we use  a visualization called a UMAP plot \citep[Uniform Manifold Approximation and Projection, ][]{McInnes2018}, described in more detail in \autoref{sec:visualization}. In \autoref{tab:awesome_table} we use these plots to illustrate how different science problems and algorithms select or classify different parts of feature space, where each point represents some piece of underlying complex data (for example an image or a spectrum).

\subsection{Types of machine learning problems}
  
The top row in \autoref{tab:awesome_table} illustrates two common types of \emph{supervised learning} problems where many labels are available, either from human labeling or additional data. These labels can be either discrete classes ({\bfseries Classification}) or continuous variables ({\bfseries Regression}). The designation of radio galaxies as either FRI or FRII \citep{Becker2021} is an example of {\bf Classification}. It requires the algorithm to recognize a wide range of morphologies, sizes, orientations and brightness distributions, and group them into two discrete classes. We indicate those using two different colors in the UMAP feature space.  When the algorithm is based on the human assignment of a continuous range of values, as in the determination of  photometric redshifts \citep{DeVicente2016}, then it is called {\bf Regression}, and can be seen as the smoothly varying color scale across feature space.

On the other hand, \emph{unsupervised learning}, illustrated in the second row of \autoref{tab:awesome_table} requires no labels. {\bfseries Clustering} can be used to break the dataset up into discrete regions in feature space, within which the objects are all close to each other, and further away from other objects. These can later be labeled by a human, such as was done in \citet{Mohale2024} on optical galaxies, to create training sets or even potentially to find new patterns such as sub-classes of sources. {\bf Anomaly detection} algorithms are used to identify those sources which are at large distances in feature space from all other sources. Such outliers in feature space were used, e.g., by \citet{Gupta2022} to locate unusual galaxies in ASKAP data.  Training data is unavailable, essentially by definition, for never-before-seen sources. 

While obtaining a large enough number of labels for supervised learning may be prohibitive for many datasets, it is often possible to provide a few strategically chosen labels to improve algorithms, illustrated in the last two rows of \autoref{tab:awesome_table}. A {\bfseries Similarity search} can be used to find more examples of a specific type of object by searching nearby feature space. This is similar to the goals of {\bf Classification}, but focused on one, or a few, specific types. The remaining three problem types in \autoref{tab:awesome_table} are all examples of \emph{active learning}, where an algorithm is iteratively improved by the introduction of a few optimally chosen labels. For instance, the RESSPECT system \citep{Kennamer2020}, uses {\bfseries Classification with active learning} to select the most useful optical transients to follow up spectroscopically in order to improve a photometric light curve classifier.

Active learning can also solve a key problem with pure anomaly detection: many anomalous sources are not scientifically interesting and must be rejected or ignored.  \astronomaly{} \citep{Lochner2021} is a general purpose {\bf Active anomaly detection} framework which combines unsupervised learning with minimal human labeling.  \astronomaly{} was successfully applied to $\sim$4 million optical galaxies in \citet{Etsebeth2024}, discovering a large number of unusual sources, including mergers, strong lens candidates and several unidentified classes.

The anomaly detection algorithms that \astronomaly{} relies on can be limited by the feature extraction method used, as alluded to in \autoref{sec:feature_extraction_intro}. This is illustrated in the bottom right section of  \autoref{tab:awesome_table} where the interesting sources occupy an extended region within feature space, rather than forming isolated patches near the edge. To solve this problem, we introduce \protegefull{}, heavily based on the algorithm from \citet{Walmsley2022}, which, instead of pure anomaly detection, leverages active learning to rapidly learn what the human trainer finds interesting. \protege{} acts, to some degree, like a \emph{recommendation engine}, similar to algorithms used by popular streaming services to recommend content to users based on their interests \citep[see][for a recent review]{Roy2022}. This form of active regression, when combined with the self-supervised features used in this paper, is highly effective at discovering truly scientifically interesting sources without requiring significant human time. 

\begin{figure*}
 \includegraphics{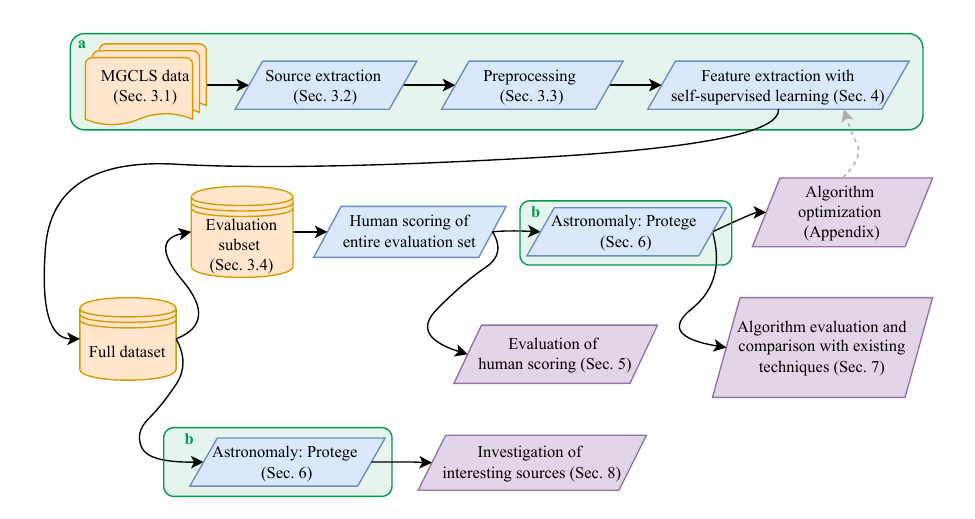}
 \caption{High level flowchart of the full methodology with references to relevant sections. The section of the flowchart indicated by $a$ outlines the procedure for extracting features from the raw data and is further expanded in \autoref{fig:flowchart_features}. After feature extraction, the analysis is divided in two, making use of a fully labeled subset of the data to evaluate performance and optimize algorithm hyperparameters and subsequently applying the methodology to the full dataset to locate and follow up high scoring sources. In both cases, \protege{} is used with the same settings and is described in more detail in \autoref{fig:flowchart_protege}.}
 \label{fig:flowchart_methodology}
\end{figure*}

\protege{}'s  goal is to identify relatively rare objects that the trainer finds scientifically interesting, but span a wide range of morphologies. It is thus broader than the {\bfseries Similarity search}, which is targeted towards specific morphologies, and also differs from {\bfseries Classification} or {\bfseries Clustering} which are also focused on finding common and related morphologies.  The only criterion for \protege{} is that the source in question is of interest to the trainer.  Its power lies in its ability to efficiently select interesting sources, as will be shown in \autoref{fig:recall}. Most exciting, however, is that \protege{} can make new discoveries, i.e.,  to find sources that the human trainer will recognize as interesting, but which do not appear to resemble those used in the training.  Examples of this will be shown in \autoref{sec:protege_finds}.

\protege{} depends on several algorithms working in concert. As we describe each of these steps, it is useful to consult the high level flow chart in \autoref{fig:flowchart_methodology}, which includes references to the relevant text sections for each functional block.

\section{Data and Preprocessing}
\label{sec:data}
\subsection{The MGCLS survey}
The MeerKAT Galaxy Cluster Legacy Survey \citep[MGCLS;][]{Knowles2022} consists of 6-10 hour, full polarization observations of 115 galaxy clusters made using the MeerKAT telescope in the L-band (900-1670 MHz). The images have a typical sensitivity of $\sim3-5$ \muJy{} and are publicly available\footnote{\url{https://doi.org/10.48479/7epd-w356}}. We make use of the high resolution total intensity images for this work, which have a typical synthesized beam size of $\sim 7.5-8 \arcsec{}$. Based on the work of \citet{Knowles2022}, we knew that this MGCLS dataset would contain a number of new and interesting sources.  Follow-up work highlighted additional unusual radio galaxies, diffuse plasmas and magnetized threads in the intracluster medium \citep{A3266,deGasperin2022, Rudnick2021, A194}.  

\subsection{Extracting Extended Sources}
\label{sec:pybdsf}
\pybdsf{} \citep{Mohan2015} is one of the most commonly used source extractors in radio astronomy. The algorithm works by first determining the rms of an image and then identifying contiguous islands of emission that are a certain threshold above the noise (typically $3\sigma$). Then, \pybdsf{} decomposes the islands into Gaussian components, which are typically used as a source catalog. Alternatively, the islands can be decomposed using shapelets to better fit extended sources.

 Automatically extracting extended sources is an active field of research and remains challenging for most source-finding algorithms (see \citet{Boyce2023} for an in-depth comparison). Often, the Gaussian components alone do not adequately describe extended sources. We propose an alternative approach by making use of the islands themselves as the input catalog, with the Gaussian  components used as a proxy for complexity. Islands that require more Gaussians are more likely to be extended sources. The position of each source is defined as the center of mass of the island, which may or may not coincide with the true visual center of the source.

As the vast majority of sources in the original dataset are unresolved or only slightly resolved, we applied a cut requiring each island to have at least four Gaussian components to be included in the catalog. Visual inspection of individual images suggested this cut included the majority of extended sources with complex morphologies.

For a more quantitative check of this procedure, we made use of the extended source catalog from \citet{Knowles2022} provided for two of the clusters: Abell 209 and Abell S295. For A209, 20 out of 33 extended sources are recovered by our method, while for AS295 19 out of 26 are recovered. Visual inspection confirms that the missed sources fall into only one of two categories. First, there were highly separated FRII galaxies where each component was described by only a single Gaussian; we did not consider such sources as ``scientifically interesting'', so their exclusion is not a problem for this experiment\footnote{Other users could well determine that these were of interest, and have to tailor their extraction parameters accordingly.}. The second category included large sources for which the human-identified center differs significantly from that detected by \pybdsf{}, so they were mis-identified, but still included. 

Examination of a larger sample of objects show that occasionally, sources that we would consider scientifically interesting can be missed by the above procedure. The star-forming ring in the Abell 548 field is a near perfect circle (see Figure 25 of \citet{Knowles2022}) and worthy of further scientific investigation.  However, it was missed because its circular nature meant \pybdsf{} could fit it with fewer than 4 Gaussian components. While the extended source extraction technique can certainly be improved, there will always be a trade-off between losing target sources and managing an increased number of contaminants. 

\subsection{Preprocessing}
\label{sec:preprocessing}
After obtaining a catalog using \pybdsf{} we then extracted image cutouts of each source. One challenge is that source size varies significantly and this property can inadvertently dominate machine learning algorithms (as found in \citet{Slijepcevic2023}). As our main interest is in radio galaxy morphology, the apparent size of the source is unimportant. We thus used the islands derived from \pybdsf{} to estimate the minimum side of a square that fully contains the source. The cutouts were then resized using \textsc{skimage} \citep{scikit-image} to a common size of 128x128 pixels. Most sources are smaller than this and so are interpolated but a few are larger than the window size and are instead downsampled.

As in \citet{Lochner2021}, we applied sigma clipping to the cutouts at $3\sigma$ to remove potentially confusing background sources. We used the package \astropy{} to estimate the standard deviation of the image and then thresholded the image at $3\sigma$. We then fit a contour around the central source using the package \opencv{} (since our sources are always in the center of the image by definition) and removed all other background sources. We investigate the efficacy of sigma clipping in \autoref{sec:appendix_hyperparams}.

Radio images tend to have high dynamic range and astronomers often resort to scaling functions, such as $log$ or $asinh$, restricting the brightness range in an image to some fraction of the maximum or a combination of these techniques to better visualize faint sources. As we have already made use of sigma clipping to remove bright background sources, we do not apply any transforms to the cutouts provided to the feature extractor. However, we do apply an $asinh$ scaling to the images viewed by the human user and do not include sigma clipping in these images. The larger context is often helpful to humans for source identification but can negatively impact performance of machine learning algorithms due to the introduction of background sources. The use of transforms is discussed further in \autoref{sec:discussion} and our choices supported by the analysis in \autoref{sec:appendix_hyperparams}.

Finally, we removed any source that results in a cut-off image due to its bounding box intersecting the edge of the original MGCLS image. This results in a final catalog of \fullcatsize{} sources.

\subsection{Evaluation subset}
\label{sec:evaluation_subset}
From this catalog, we created a fully labeled subset of sources. We use this evaluation subset to compare \protege{} against other approaches in \autoref{sec:evaluation}. In the process of evaluating performance, we also used this subset to explore the impact of hyperparameters for both \protege{} and traditional \astronomaly{} (\autoref{sec:hyperparameters} and \autoref{sec:appendix_hyperparams}).

\begin{deluxetable}{ccc}

 \tablehead{\colhead{Score} & \colhead{Example sources} & \colhead{Number of sources}}
 \tablecaption{Human scores used for the evaluation subset. \label{tab:classes}}
 
  \startdata
 1 & Unresolved, artefacts & 384\\
2 & ``Normal'' FRI or FRII & 439\\
3 & WAT, double-double, disturbed & 122\\
4 & NAT, S-shaped, spiral, diffuse & 62\\
5 & X-shaped, turbulent, unknown & 24\\
\enddata

\end{deluxetable}

\begin{figure*}
 \includegraphics{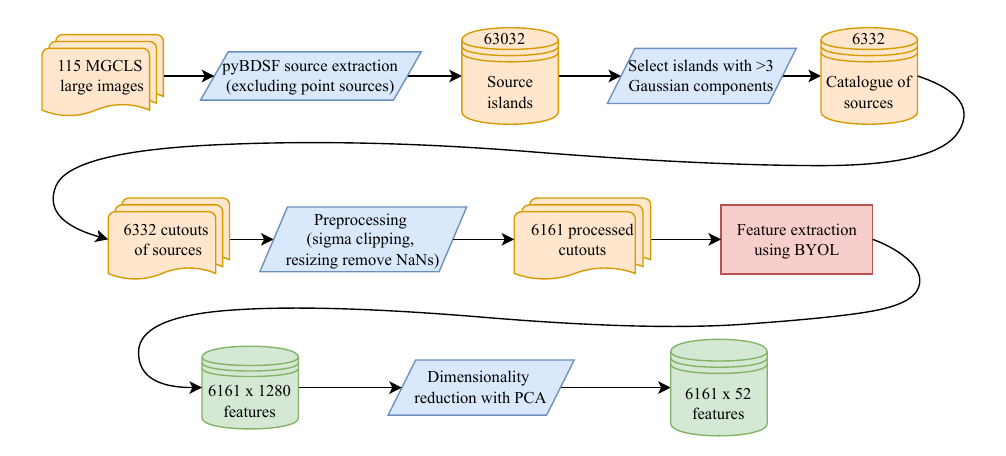}
 \caption{Flowchart of the preprocessing and feature extraction process.}
 \label{fig:flowchart_features}
\end{figure*}

To create the evaluation subset, we selected the two fields that have an extended source catalog, (Abell 209 and S295), and then 18 other randomly selected fields in the MGCLS. After source extraction and preprocessing, this results in a catalog of \evaluationcatsize{} sources, which were then scored by author ML on a scale of 1 to 5. \autoref{sec:interestingness} discusses further what is meant by an interesting source and the subjective scoring procedure used throughout this paper. \autoref{tab:classes} shows a break down of the number of sources per score and a rough description of the types of sources corresponding to that score. We define any object with a score of 4 or 5 to be of high interest, which corresponds to $8.34\%$ of the evaluation subset. The evaluation subset human scores are included as a separate column in the \href{\fileslink}{catalog} provided with this paper.

\section{Self-supervised Learning for Feature Extraction}
\label{sec:features}

After preprocessing, the next step is to simplify the images in a process known as feature extraction. Features should be low dimensional but still capture key information in the images to ensure successful application of the downstream algorithm. The choice of feature extractor can dictate the types of sources the algorithm will be sensitive to.

\citet{Lochner2021} used a simple approach to feature extraction which focused on shape parameters. Although initially designed for optical galaxies, this works remarkably well for radio galaxies (see \autoref{sec:evaluation}). However, this  simple approach is insufficient for our more ambitious goal of using active learning to group sources of interest in feature space, and so we make use of a more sophisticated feature extraction scheme. 

The deep learning algorithm Bootstrap Your Own Latent \citep[BYOL,][]{Grill2020} has emerged as a powerful method to automatically learn a low dimensional feature space such that similar looking objects should have similar features. \citet{Mohale2024} showed that BYOL can be used to extract meaningful representations of both optical and radio galaxies. These features were used to automatically cluster together galaxies with similar morphology, as well as to detect anomalous galaxy types such as mergers. We used the exact same technique in this work to extract features for the MGCLS sources.

The full procedure required to extract features from the MGCLS images is fairly complex, involving many steps. \autoref{fig:flowchart_features} shows a flowchart of the process, including the preprocessing steps described in \autoref{sec:data}, as well as the feature extraction methodology described in the sections below.

\subsection{BYOL}
\label{sec:byol}
BYOL is a self-supervised learning technique that aims to automatically learn useful features for a dataset without requiring any labels for the data. BYOL operates on the key insight that if an image is \emph{augmented}, in other words changed in some trivial way such as applying rotation, zooming, blurring etc., the features that describe that image should not change significantly. The goal is therefore to train a neural network to predict features such that augmented pairs produce similar values.

BYOL consists of a pair of identical neural networks: an online and a target network. The goal of the online network is, given an augmented view of an image, to predict the output of the target network which has a different augmented view of the same image. The two networks learn in such a way that they should converge to the same representation of the data. The second-to-last layer of weights of the online network are used as features.

Any network architecture can be used for the online and target networks. For this work, we use EfficientNet-B0 \citep{Tan2019} due to its lightweight architecture which makes it computationally efficient and suitable to the relatively small dataset size.\footnote{While investigating other architectures would be useful, the high computational cost made such an analysis prohibitive for this work.} \citet{Mohale2024} showed that initializing the network with weights learned from training on the terrestrial dataset ImageNet \citep{ImageNet}, instead of random weights as is commonly used, results in much faster convergence and better performance. We thus initialize the EfficientNet network with these weights and fine-tune the network to our data using the packages \textsc{pytorch} \citep{pytorch} and \textsc{byol-pytorch} \citep{chen2020}.

It is the second last layer of EfficientNet, called \texttt{avgpool}, that is used as features. In normal supervised learning the last layer would be a final fully connected layer which aims to predict the class or quantity the network is being trained to predict. For self-supervised learning, which aims to extract useful features, this layer is ultimately discarded, and the output of the preceding layer is used. For EfficientNet-B0, this results in 1280 features.

\subsection{Hyperparameters}
\label{sec:hyperparameters}
BYOL has several hyperparameters and training choices that can impact performance. In addition, the choice of augmentation strategy is critical for good results and is expected to be different for astronomical data compared to those used for terrestrial image datasets. Fortunately, \citet{Slijepcevic2023} performed an extensive study of these choices on a radio astronomy dataset. Although \citet{Slijepcevic2023} were primarily interested in classification performance, it is reasonable to expect similar choices will work well for general feature extraction and other downstream tasks. We thus do not perform an extensive analysis of hyperparameters and augmentation choices, and rather use a similar training procedure as \citet{Slijepcevic2023}, highlighting variations where appropriate. \autoref{sec:appendix_hyperparams} has a detailed discussion on the impact of hyperparameter choices, using the evaluation subset (\autoref{sec:evaluation_subset}).

We use Adam \citep{Kingma2017} as optimizer, with a batch size of 32 and a base learning rate of $0.0005$ (which is scaled by batch size/256). We find that training for just 100 epochs is sufficient for excellent features (see \autoref{fig:loss} in the Appendix). We use almost identical augmentations as \citet{Slijepcevic2023}, as shown in \autoref{tab:augmentation}. The only difference is the size of the center crop, since our images are rescaled such that the source should occupy most of the cutout.

\subsection{Dimensionality reduction}
\label{sec:pca}
As described in \autoref{sec:byol}, the number of features from the network is 1280. This is generally too high dimensional for most downstream applications so must be reduced. While there are many approaches to this problem, a common solution is Principal Component Analysis \citep[PCA, ][]{Hotelling1933}. PCA decomposes a dataset into a new coordinate system such that each orthogonal component aligns with directions of maximal variance. After the decomposition, the majority of components can be discarded, retaining only those that describe the most variance in the dataset - the principal components. Dramatic reductions in dimensionality are possible while losing only a small amount of information. We apply a threshold of 95\% variance, reducing the original 1280 features to just 52 for the full dataset.

\subsection{Visualization}
\label{sec:visualization}
For unsupervised learning tasks, it is useful to visualize the extracted feature space to gain intuition about the effectiveness of the method and where classes of objects may lie in relation to each other. Naturally this is very challenging with a high dimensional space. As is standard practice in the field, we make use of the algorithm Uniform Manifold Approximation and Projection \citep[UMAP,][]{McInnes2018}, implemented using the \textsc{umap-learn} python package \citep{Sainburg2021}. The aim of visualization algorithms like UMAP is to embed the high-dimensional feature space into a two-dimensional one, while preserving relative distances between data points. This means that after extracting features from the images, reducing them using PCA and further embedding them using UMAP, similar looking galaxies \emph{should} appear close together in the UMAP plot.

UMAP achieves this by constructing a fuzzy topological representation of the features and then learning a lower dimensional approximation to this via an optimization algorithm. While UMAP plots can be useful tools for visualization and can actually work on high dimensional spaces, the algorithm is not usually used as a dimensionality reduction algorithm directly. This is because a linear algorithm like PCA better preserves global structure than embedding algorithms like UMAP, which can occasionally create artificial clusters during the embedding process. This can be mitigated to some degree by careful choice of hyperparameters. Following \citet{Mohale2024}, we set the number of neighbors to 15 and the minimum distance to 0.01 for all UMAP plots.

\section{What makes a source interesting?}
\label{sec:interestingness}

Since our goal is to identify scientifically interesting sources, i.e., those worthy of further study (hereinafter \emph{interesting}), it is worth considering what that really means in practice. In the simplest terms, and with no intent to be frivolous, an \emph{interesting} source is whatever an individual researcher or community decides it is.  This definition is not trivial because it strongly influences  awards of telescope time, funding proposals, reviews of journal articles, and the focus of theoretical and simulation work. In the case of large surveys, defining what is \emph{interesting} is imperative, so that limited time and resources can be focused on the most promising sources and scientific questions. Since objects of scientific interest will differ between different groups and over time, it is not useful to search for algorithms to find ``objectively'' interesting sources, only those that are relevant to the current effort.

There are currently two complementary approaches to this challenge.  First is to explore algorithms that select samples of sources from a larger group, and to then judge how well each algorithm has preferentially selected \emph{interesting} sources.  The use of image complexity \citep{Segal2019, Segal2023}, is one such successful method. The second approach is to pre-define what is \emph{interesting}, in terms of identifying sources with a specific morphology, and then use \emph{similarity} algorithms  \citep[e.g.][]{Walmsley2022} to extract them from a large data set.

An intermediate approach is used by \astronomaly{} \citep{Lochner2021}, which combines anomaly detection with some active learning based on human input to reject sources the user identifies as not \emph{interesting}. However, the combination of traditional anomaly detection and subsequent active learning cannot handle the situation where \emph{interesting} sources are buried in feature space (see \autoref{sec:ml_intro}).

\protege{} is intended to take the approach of \astronomaly{} to a different level -- to recognize the broad range of morphologies that its scientist trainer finds \emph{interesting}, and then select those rare cases from the enormous data sets produced in modern surveys. This breadth in morphology is the essential challenge -- can a machine find similarities in feature space that allow it to mirror the interests of its trainer? Can it go even further and find sources with completely different morphologies from those in its training, that are also \emph{interesting}? This is the challenge for which \protege{}, described in \autoref{sec:anomaly}, is designed to address.

\begin{figure}
 \includegraphics[width=\linewidth]{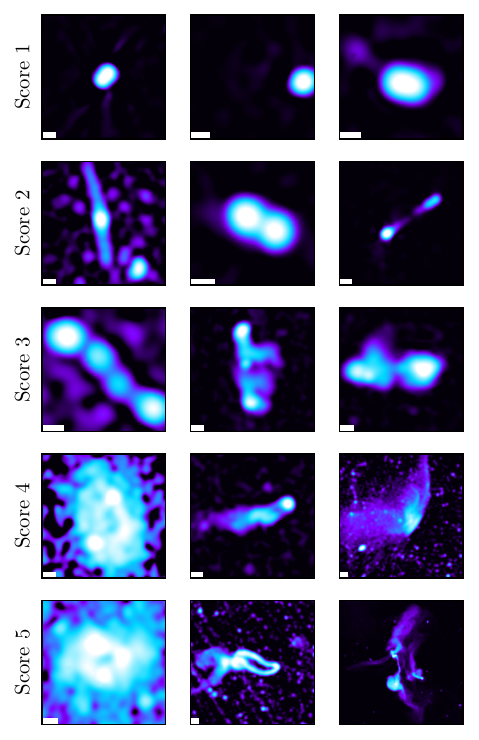}
 \caption{Examples of sources with their human-assigned score. Each image has been normalized using an $asinh$  scaling with a threshold at 90\% of the maximum brightness to highlight faint features. The beam size for each image ($\sim7.6\arcsec$) is indicated with a bar in the lower left. }
 \label{fig:examples}
\end{figure}

 To accomplish this, the trainer \citep[or ``oracle'', as used in the active learning literature][]{ActiveLearning2023}, provides input by assigning a numerical score of 1-5 on how \emph{interesting} each of a given group of sources appears. The sources are therefore not labeled, but given a numerical score which can be used by the algorithm. \autoref{fig:examples}, shows example sources from this work, along with their scores.

\begin{figure}
 \includegraphics{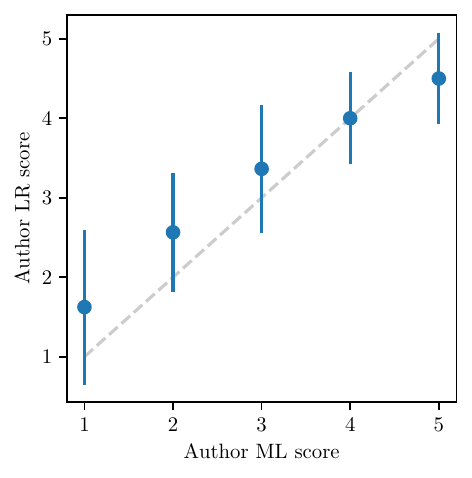}
 \caption{Comparison between the two lead authors' scoring for a random sample of 100 sources. The error bars represent the standard deviation of author LR's scores for a given value of ML's scores.}
 \label{fig:comparison_larry_michelle}
\end{figure}

Independent of the success of the algorithm, in order  to be scientifically useful, there must also be some agreement on \emph{interesting} sources among multiple users. To investigate the sense to which this is true in the current case, we performed a small experiment comparing the scores of the two authors of this paper, ML and LR.  ML is relatively new to the field of radio galaxies while LR has decades of experience.  They had several conversations, on different projects, of what sources appeared most \emph{interesting}. Thus, there was some common understanding, although not a prescription for finding specific morphologies. They each assigned a score to a random subset of 100 sources, and their scores were compared as shown in \autoref{fig:comparison_larry_michelle}.  Perfect agreement is neither expected (see \citet{Savary2022} and \citet{Rojas2023} for differences in expert labeling of strong lens candidates), nor necessarily desirable. We found a reasonable agreement between the two scorers, with a non-parametric Pearson correlation coefficient of 0.70, and only 12\% of the source scores differing by more than one point. This gives us confidence that there is sufficient agreement on what constitutes an \emph{interesting} source to make the approach meaningful.

\begin{figure}
\centering
 \includegraphics{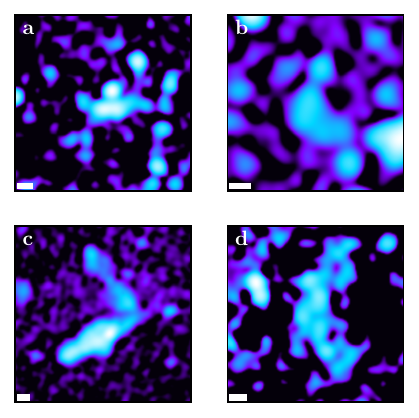}
 \caption{Examples of sources author LR scored high while ML scored low. The images have been normalized with $asinh$ scaling.}
 \label{fig:examples_L_high_M_low}
\end{figure}

\begin{figure}
\centering
 \includegraphics{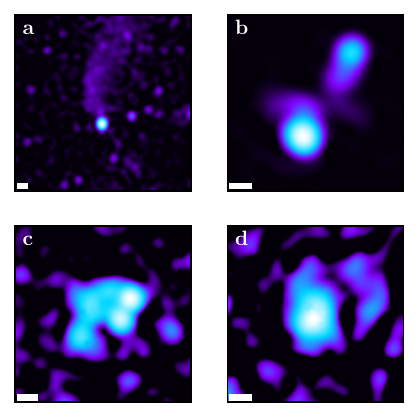}
 \caption{Examples of sources author ML scored high while LR scored low. The images have been normalized with $asinh$ scaling.}
 \label{fig:examples_L_low_M_high}
\end{figure}

However, it is also instructive to explore where there was significant disagreement between the two scorers, as shown in \autoref{fig:examples_L_high_M_low}  and \autoref{fig:examples_L_low_M_high}.  Such disagreements form  an excellent basis for scientific discussion, as the community seeks to reach a common understanding.  Although not pursued further here, it would be interesting to have a variety of different users each train \protege{}, and compare the resulting samples. 
Another opportunity, as discussed below, is to think about \protege{}'s trained scoring as an oracle in its own right, and look for novel examples of what may be \emph{interesting} to the human users, that they may have missed.

\newpage
\section{Finding \emph{interesting} sources}
\label{sec:anomaly}

\begin{figure*}[ht!]
 \includegraphics{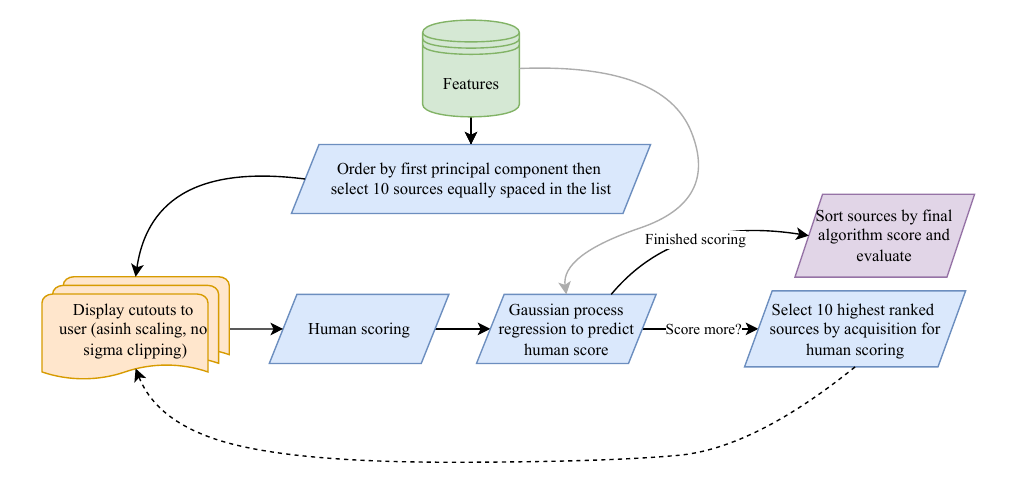}
 \caption{Flowchart of \protege{}.}
 \label{fig:flowchart_protege}
\end{figure*}

\astronomaly{}'s flexibility as a general framework makes it ideal for application on a novel and unlabeled dataset like MGCLS. By relying on an initial feature extraction step, \astronomaly{} can work with essentially any type of astronomical data. We use the extracted features from a trained BYOL model, as described in \autoref{sec:features}, with the aim of finding sources that we, the authors, find \emph{interesting}.

As discussed in \autoref{sec:ml_intro}, \citet{Walmsley2022} found that for deep features, the traditional anomaly detection algorithms used in \citet{Lochner2021} did not perform well. They effectively bypassed anomaly detection entirely and used a regression algorithm, specifically Gaussian processes \citep{Rasmussen2006}, coupled with a traditional active learning approach, to rapidly locate the objects of interest to the user.

While we did explore the use of traditional anomaly detection algorithms in this work (\autoref{sec:evaluation}), we found that the approach developed in \citet{Walmsley2022} has the highest performance for this dataset and so is what we will focus on. Because this approach is philosophically different from traditional anomaly detection, we give it a new name to avoid confusion, including it as a module within \astronomaly{}. Since the method aims to learn what is \emph{interesting} to the user (not strictly speaking anomalous), and does this by iteratively requesting instruction on what sources are \emph{interesting}, we name it \protege{}.

It is worth reiterating that, unlike anomaly detection algorithms, \protege{} can be trained to find any source of interest, even if it is common (for example, FRII radio galaxies). However, for a class that has many examples, a trained supervised learning algorithm will generally outperform \protege{}. The goal of \protege{} is rather to act like a recommendation engine for new discoveries, fine-tuning its search for sources a user may find \emph{interesting} based on that user's input.

\subsection{\protege{}}
\label{sec:protege}

\begin{figure*}[htp!]
 \includegraphics[width=\textwidth]{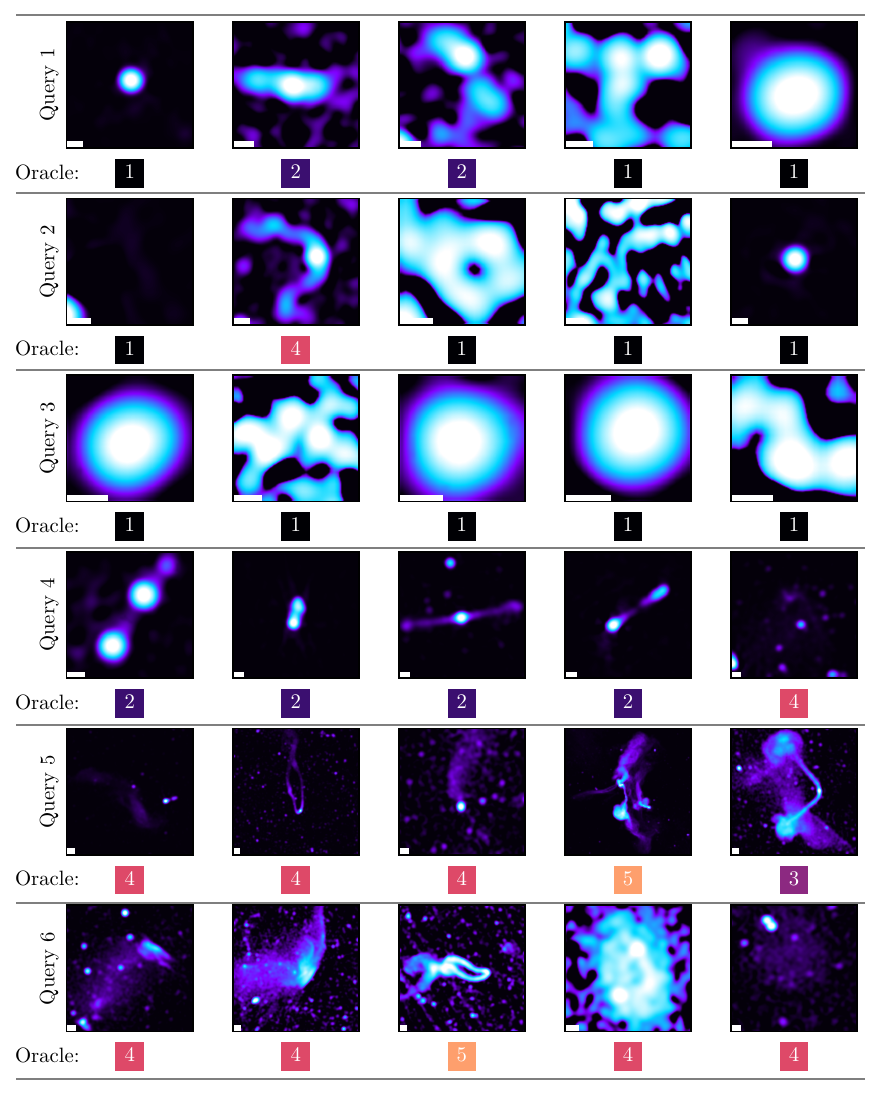}
 \caption{Demonstration of the interactive training procedure. In a ``query'', a set of objects is shown to the ``oracle'' (the user) who provides an interest level score for each object (1 for compact sources or pure noise to 5 for highly unusual morphology). Images are normalized using $asinh$ scaling with the brightness truncated to 90\% of maximum. The active learning algorithm described in \autoref{sec:anomaly} is then applied, the objects are sorted by acquisition score (which prioritizes which objects will provide the algorithm with the most useful information), and the next set of objects are queried. Within just a few iterations, the algorithm hones in on high human-scoring sources. }
 \label{fig:oracle}
\end{figure*}

\protege{} essentially aims to learn a function across feature space that predicts what score a user will give each source. As with other active learning algorithms, it uses an acquisition function to prioritize which sources, if scored by the trainer, would be most informative in training the algorithm. Different active learning approaches use different acquisition functions.

\autoref{fig:flowchart_protege} shows a flowchart broadly outlining the algorithm. The first step is to select an initial set of sources to query. These will be shown to the user (or ``oracle''), using \astronomaly{}'s front end interface, to be scored from 1 to 5 based on how \emph{interesting} the user thinks the source is. While the range of scores is arbitrary, a breakdown of 5 levels is sufficient to allow a clear gradient for an algorithm to learn.

 The goal of the initial set of sources presented in the queries for human scoring is to broadly sample the full feature space. We take advantage of PCA by selecting the ten sources equally spaced from a list ordered by the first component (i.e. the first feature). This differs from the approach in \citet{Walmsley2022} where the initial sample was selected randomly. We found a PCA-based approach higher performing and more consistent when the training was repeated.

 In the next step, \protege{} predicts the score of the entire dataset based on the training data provided so far, and selects the next set for human scoring. We use a Gaussian process (GP) for predicting the user score because it provides an estimate of the uncertainty, which can then be used to calculate the acquisition function. A Gaussian process effectively models the data using a Gaussian probability distribution in function space, predicting both the mean and standard deviation of the target quantity.  We use the same acquisition function used in \citet{Walmsley2022}: the \emph{expected improvement} \citep[EI,][]{Mockus1991, Jones1998}. This function, described in detail in \citet{Walmsley2022}, includes a hyperparameter $\epsilon$ which can be used to tune the trade-off between exploitation (low values of $\epsilon$) and exploration (high values of $\epsilon$). 
 We used a value of $\epsilon = 3$ (half way between our minimum and maximum expected user scores), which encourages the algorithm to explore regions of high uncertainty. This makes it less likely to get fixated on a particular region, and increasing the chances of new discoveries. An analysis described in \autoref{sec:appendix_hyperparams} showed that as long as this parameter was higher than 0.5, it did not make a significant difference to which sources were discovered.

After predicting the user scores for the entire dataset and calculating the acquisition function, \protege{} then selects the ten sources with the highest values of the acquisition function and presents them as the next query. While any number of sources can be scored in each iteration, we find that ten is a good balance between providing a reasonable amount of training data for the GP and updating rapidly enough to locate high human-scored sources quickly.

This process is repeated until the trainer determines that the algorithm is trained well enough and is presenting many more sources of interest to them than uninteresting sources. The list of sources can then be sorted by the most recent predicted score from the GP and is now ready for user inspection. \autoref{fig:oracle} illustrates an example of the interactive approach of \protege{}. \autoref{sec:appendix_regression} investigates the performance of the GP algorithm in predicting user scores.

\section{Performance evaluation}
\label{sec:evaluation}
We use the evaluation subset described in \autoref{sec:evaluation_subset} to compare our method with alternative approaches. The goal was to determine whether \protege{}, traditional \astronomaly{} or a non-machine learning approach would best recover \emph{interesting} sources with minimal human labeling. 

\subsection{Complexity comparison}
\label{sec:complexity}
Several recent works \citep{Segal2019, Segal2023} have shown that anomalies in radio images, and thus sources which might be considered \emph{interesting},  can be effectively detected using measures of complexity. Here, we use a simplified measure of complexity, described below, as a demonstration of what can be achieved without additional processing steps. 

The number of Gaussian components in each island identified with \pybdsf{} was taken as a measure of source complexity. Sources with the most components are given higher scores, extending the idea behind the minimum four-component cut used to create the catalog (\autoref{sec:pybdsf}). This simple procedure provides a useful baseline to judge whether the machine learning approach of \protege{}  is sufficiently superior to a simple catalog ordering to warrant the additional computational cost.

Our simple complexity measure does not, however, provide features that can be used by active learning algorithms to target specific morphologies and cannot be used for downstream tasks such as disregarding artefacts, applying classification or similarity searches.

\subsection{Ellipse-based feature comparison}
\label{sec:ellipses}

In \citet{Lochner2021}, \astronomaly{} originally used hand-crafted morphology features based on fitting ellipses to isophote contours. This was followed by a traditional anomaly detection algorithm \citep[isolation forest,][]{Liu2008} combined with a active learning. Although originally applied to optical galaxies, it is reasonable to assume a similar approach would work for radio sources so it was the first approach we used. However, inspired by the high performance of deep learning based-feature extraction used in \citet{Walmsley2022}, we moved on to the methodology described in \autoref{sec:features}. The ellipse-based approach, described in more detail in \autoref{sec:appendix_ellipses} is included here only as another baseline against which to compare. As there is some randomness in the isolation forest algorithm, we run the full \astronomaly{} pipeline 10 times to estimate the variance in performance on the evaluation subset.

\subsection{BYOL-based feature comparison}
We compared the previous two approaches to our new method, BYOL-based feature extraction (\autoref{sec:byol}) combined with \protege{} (\autoref{sec:protege}). 

One advantage of a self-supervised algorithm like BYOL is that a larger, unlabeled dataset can be used to train the algorithm, improving the features extracted. To investigate the effect of training set size on performance, we considered three different subsets:

\begin{itemize}
 \item The evaluation subset (\evaluationcatsize{} sources).
 \item A random subset of \randomsubset{} sources, {\bfseries excluding} any source in the evaluation subset.
 \item The full dataset (\fullcatsize{} sources).
\end{itemize}

Since the augmentations BYOL relies on are applied randomly, we ran trained BYOL 10 times on each subset allowing an estimate of the model variance.

BYOL trains by minimising the loss - the mean-squared error between the online and target representations. \autoref{fig:loss} in the Appendix shows the loss as a function of epoch for each of these three subsets. Not surprisingly, increasing the amount of data improves the loss significantly. 

Each iteration produces a different feature set, trained using the same subset of data but different random augmentations. The features for the evaluation subset are then passed to \protege{} and its performance evaluated. \protege{} is trained using the procedure outlined in \autoref{sec:protege}, in batches of 10 until 100 sources have been given human scores.

\autoref{sec:appendix_hyperparams} explores using the original \astronomaly{} approach with BYOL features, which performed poorly, so we exclude it here.

\subsection{Results from the Evaluation Subset}
\label{sec:results}

\begin{figure}
 \includegraphics{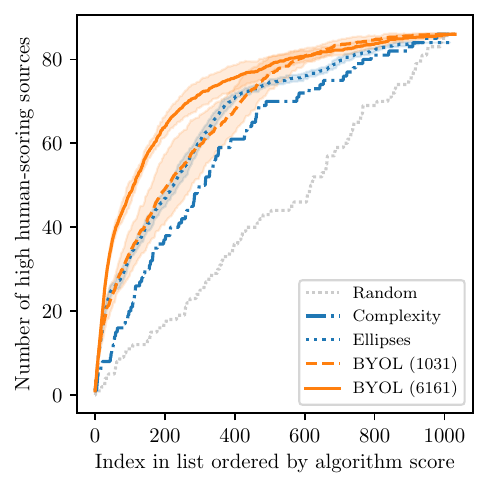}
 \caption{Number of sources author ML scored 4 or 5 in the evaluation subset when ordered by the predicted score from several algorithms: random ordering (\emph{Random}), ordering by number of components in an island from \pybdsf{} (\emph{Complexity}), ordering by \astronomaly{} using ellipse-based features with an anomaly detection algorithm and active learning (\emph{Ellipses}), and two cases of ordering using the BYOL features combined with \protege{} for active learning (\emph{BYOL}). The numbers in brackets for BYOL refer to the number of sources used to train the algorithm. The 3500 source subset is excluded due to the curve's similarity to the full dataset curve. The envelope around the lines indicate the standard deviation from 10 repetitions. As it results in the steepest curve, BYOL combined with \protege{} outperforms all other methods when sufficient training data is used.}
 \label{fig:recall}
\end{figure}

\autoref{fig:recall} shows the recall as a function of index in an ordered dataset. In other words, for any given number of sources ($N$ on the x-axis) starting from the highest ranked, the y-axis reports the number of those sources with a human score of 4 or 5. The steeper this plot, the better the algorithm is at recovering \emph{interesting} sources, without needing a user to look through a vast number of objects. A random ordering of sources results in a linear growth in the number of sources with human scores of 4 or 5, as expected, up to the full number (86) of such sources.  Ordering by our complexity proxy performs surprisingly well, indicating that such methods have merit in improving the efficiency of identifying sources of interest. 

The ellipse-based method described in \autoref{sec:ellipses} also performs well, despite being developed for optical galaxies. It is actually comparable to the algorithm using BYOL features combined with \protege{} in our initial training experiment. However,  BYOL + \protege{}  improve significantly as the BYOL algorithm is trained on more data. The random sample of \randomsubset{} was excluded from the plot for clarity, as the curve is almost identical to that of the full dataset. This implies that these results are not due to ``data leakage'', as BYOL was trained on a subset excluding the evaluation subset. However, the performance improves slightly and the variance decreases when using all available data. We thus expect the algorithm to perform best on the full dataset when BYOL is also trained using the full dataset. Note that there is no way to similarly scale the ellipse-based method.

\begin{figure*}
 \gridline{\fig{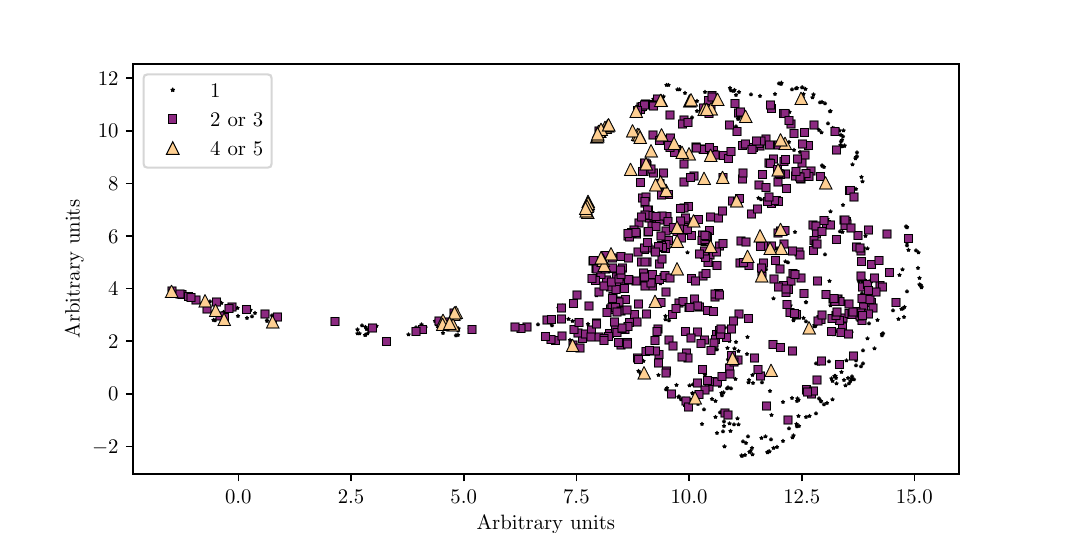}{1\textwidth}{}}
 \gridline{\fig{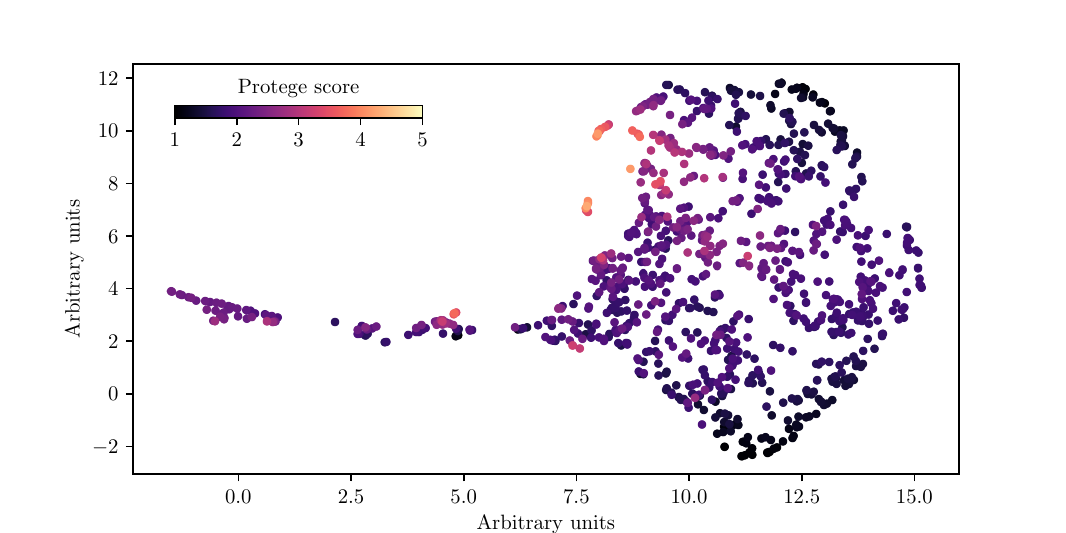}{1\textwidth}{}}
 \caption{UMAP visualization of the BYOL feature space. Each point represents a distinct source in the evaluation subset. \emph{Upper panel:} points are colored by their underlying human score, showing that the sources with a score of 1 collect towards the edge of feature space while the sources with a score of 4 and 5 form clusters embedded throughout feature space. \emph{Lower panel:} the same sources colored by the predicted \protege{} score, after training on 100 labels selected by the acquisition function. It can be seen that, even with only a few labels, the algorithm learns which regions of feature space to focus on, although naturally a few high human-scoring sources are missed.}
 \label{fig:umap_evaluation}
\end{figure*}

To make the practical application of \autoref{fig:recall} clearer, we consider a simple thought experiment. If a user were willing to look through all \evaluationcatsize{} sources,  they would naturally find all the high-scoring sources. However, looking through all the sources is already tedious and will quickly become impossible as dataset sizes increase. If instead, a user only had the time and energy to view 100 sources, they would discover only 8 high-scoring sources if the sources were chosen at random, reaching 20  using a list ranked by complexity measure or 32 using traditional \astronomaly{} with ellipse features.  But using \protege{} with BYOL, an impressive 56\% (48 out of 86) sources were found, even though less than 10\% (100/\evaluationcatsize{}) of the sources had been checked. Given that the original full catalog consists of \originalcatsize{} sources, the chance of finding a high human-scoring source looking through just 100 examples from it is extremely small ($\sim$0.084 in a random sample of 100).

\autoref{fig:umap_evaluation} shows two UMAP visualizations (described in \autoref{sec:visualization}) of the BYOL features (using one of the full dataset runs) for the evaluation subset. The top panel reveals that ``boring'' sources (given a human score of 1) are grouped together near the edges, making it very easy for \protege{} to discard them. There are also fairly pure regions containing sources of ``medium interest'' (human scores of 2 or 3) which can be ignored. While some sources with a human score of 4 or 5 are embedded among others, most are found clustered in a few key regions. These correspond to the high scoring regions in the lower panel, where points are colored by \protege{} score. These plots demonstrate two key points: BYOL is effective, but not perfect, at grouping together sources the authors find \emph{interesting} and \protege{} is able to find the regions where these sources tend to reside, even with only 100 training labels provided.

\subsection{Sources \protege{} finds, but humans missed}
\label{sec:protege_finds}

\begin{figure}
\centering
 \includegraphics{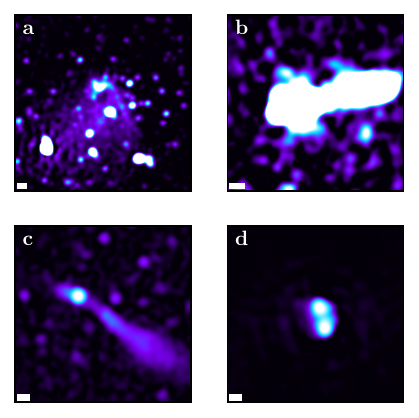}
 \caption{Four examples of sources given a high score by \protege{} and a low score by humans. Upon reinspection, we agree that the sources in $a$, $b$, and $c$ are worth further consideration,  as described in the text, but not $d$. All images are produced with asinh scaling. To reveal faint structure, the maximum brightness is truncated to 10\%, 10\%, 90\% and 90\% of the maximum for panel $a$, $b$, $c$ and $d$ respectively.}
 \label{fig:protege_finds}
\end{figure}

A key question to explore is whether \protege{} can find sources which are \emph{interesting}, but were initially missed by humans. \autoref{fig:protege_finds} shows four examples that were scored high by \protege{} but low by the authors. Some of these sources contain subtle diffuse structures that were missed, which can happen when viewing large samples of images. Source \emph{a} in \autoref{fig:protege_finds} contains a faint, diffuse structure (halo), located in the center of cluster Abell S295, with a remarkable long filament. Source \emph{b} is a radio galaxy that appears typical at first glance, but closer inspection reveals transverse filamentary structures similar to the ``ribs'' described in \citet{Rudnick2021}. The asymmetry of source \emph{c}  warrants a second look from a science perspective. 

These examples demonstrate the power of machine learning to not only locate rare sources in large datasets, but also to locate sources that may have been missed by humans and deserve a second look. The less biased nature of machine learning algorithms, coupled with their ability to detect subtle structures not initially obvious to human users, provides a powerful tool for scientific discovery.

We also note that the unusual and scientifically interesting source SAURON (originally detected using the ellipse-based method) appears second in the list when ordered by \protege{} score (using BYOL trained on the full dataset as feature extractor), placing it in the top 0.2\% of the data. 

The current algorithm is not perfect. Source \emph{d} \autoref{fig:protege_finds} is a high-scoring source which we considered uninteresting, even on a second look. It is unclear why this source appears in the same area of feature space as the high human-scoring sources. A similarity search (\autoref{sec:ml_intro}) revealed that all nearby sources indeed look similar and would also be given a low human score. This is an indication of an imperfect feature space, which could perhaps be improved with more training data and better preprocessing. However, there were very few examples of such sources (in fact, it was a challenge to find one) and we agreed with the majority of \protege{}'s top recommendations.

\subsection{Sources with high human but low \protege{} scores}

\begin{figure}
 \includegraphics{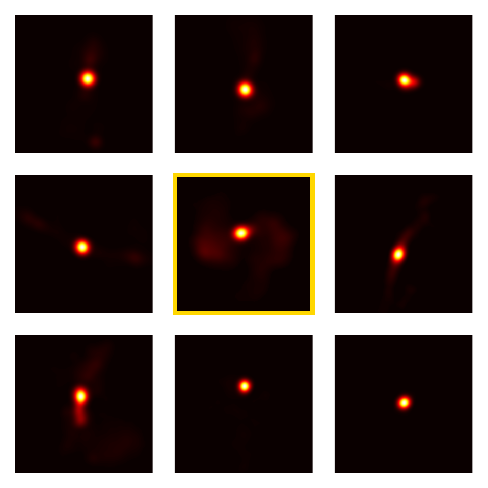}
 \caption{Similarity search of feature space for a high human-scoring source for which \protege{} gave a low score. The source of interest is displayed in the center with its eight nearest neighbors (by Euclidean distance) around it. These images have not been rescaled and are identical to the input to the feature extraction algorithm. Viewing the central source with $asinh$ scaling reveals the faint structure around the central core, so this source was given a low score purely because of the preprocessing scaling. Nearly all high human-scoring sources that were missed by \protege{} fall into this category.}
 \label{fig:protege_fails_1}
\end{figure}

\begin{figure}
 \includegraphics{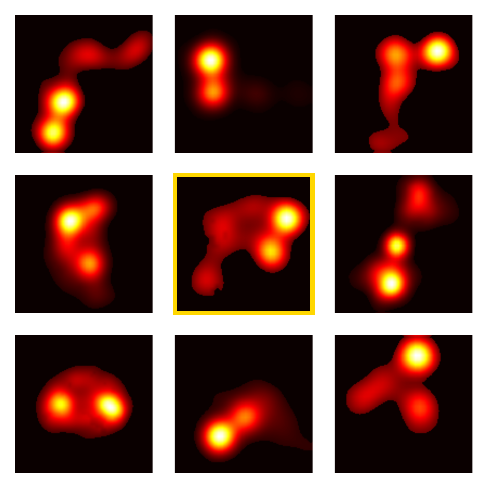}
 \caption{Similarity search of feature space for a second high human-scoring source (center) for which \protege{} gave a low score, showing its eight nearest neighbors. This source is similar to other double core sources so it's difficult to see how the algorithm could have distinguished between them.}
 \label{fig:protege_fails_2}
\end{figure}

\autoref{fig:umap_evaluation} shows that some sources with high human scores are missed by \protege{}, primarily because they end up with similar features to sources the authors were not interested in. A closer look at a high-human scoring source which \protege{} ranked low can be seen in \autoref{fig:protege_fails_1}, with its eight nearest neighbors, found using a similarity search. This source has faint diffuse emission which is largely invisible with the linear scaling used by \protege{},  but obvious with \emph{asinh} scaling used for re-inspection. Thus it is the preprocessing which is ineffective. The majority of ``missed'' sources fall into this category, indicating that improving the preprocessing could be key to improving the overall method (see \autoref{sec:discussion} for further discussion).

A similarity search for the source in \autoref{fig:protege_fails_2} shows that its neighbors have a similar morphology to other sources which are mostly slightly resolved. This is one of only a few examples where the source of interest is simply not very well-resolved and easy to mistake for other types.
 
 On balance, the overall high efficiency of \protege{} in identifying the most \emph{interesting} sources (\autoref{fig:recall}), and the discoveries that it makes that are missed by humans, show that it is a very important tool as datasets become extremely large.  At the same time, it is important to acknowledge that it is imperfect, missing some sources that humans would rank as \emph{interesting}.

\section{Recommendations from \protege{}}
\label{sec:full_dataset}

\begin{figure*}
 \includegraphics{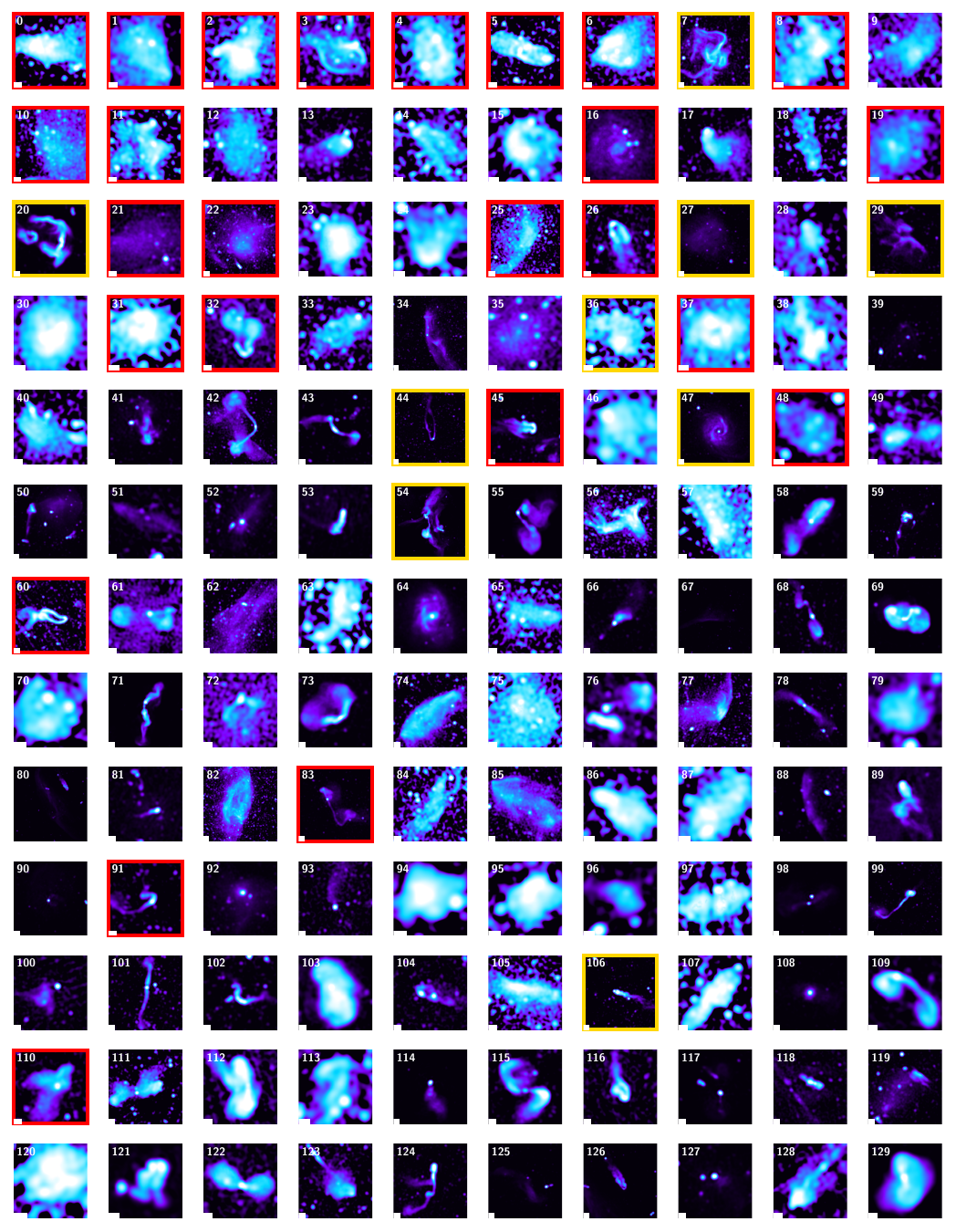}
 \caption{Highest ranked sources according to \protege{}. Images have been normalized using an $asinh$  scaling with a threshold at 90\% and the beam size ($\sim7.6\arcsec$) is indicated with a bar in the lower left. Known sources that have published figures from \citet{Knowles2022} and related literature are highlighted with a yellow border, while new recommendations discussed in the text have a red border. Figure continued on next page.}
 \label{fig:top_scores}
\end{figure*}

\begin{figure*}
 \includegraphics{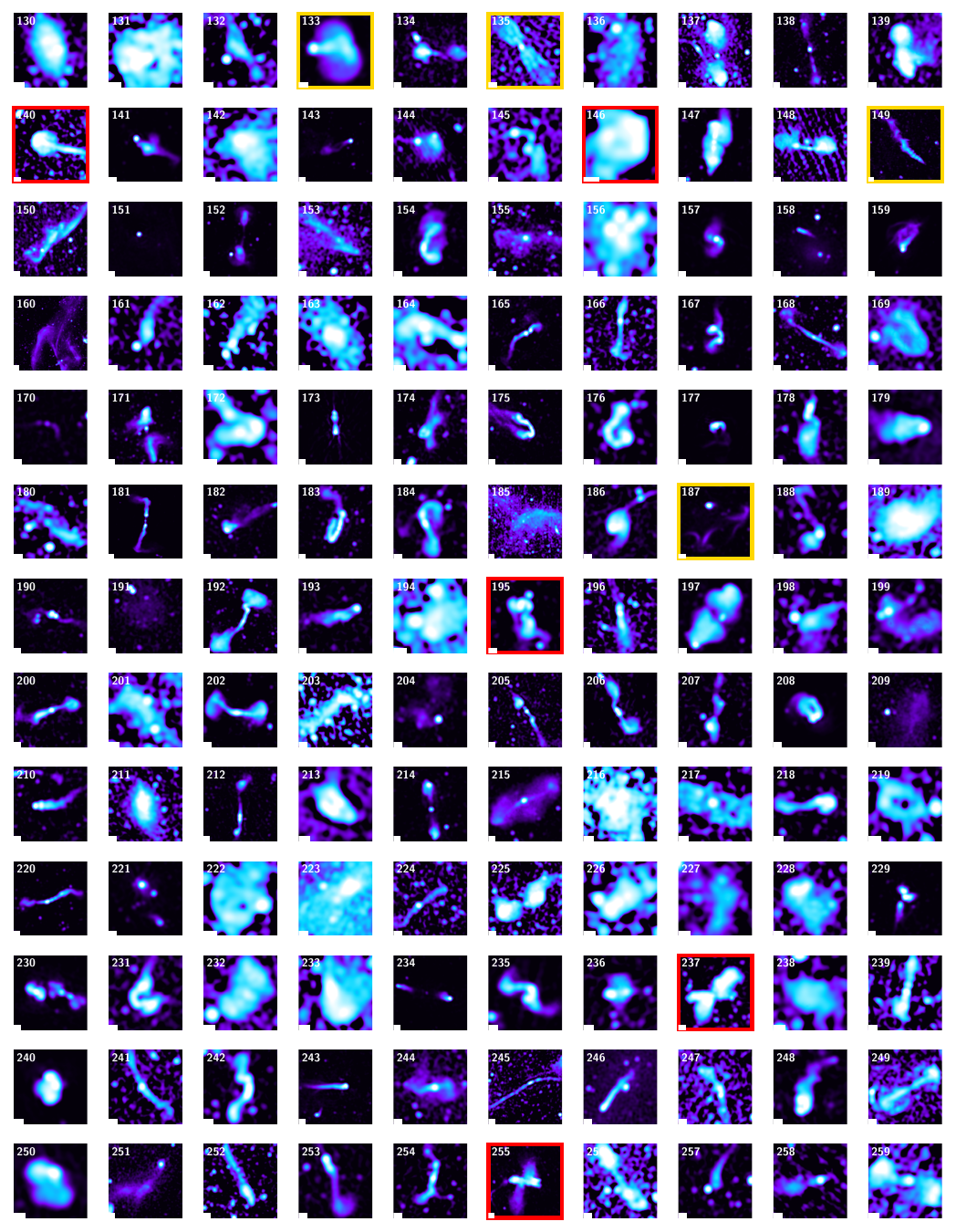}
\end{figure*}

So far, we have demonstrated, on a sample of \evaluationcatsize{} sources,  how \protege{} can be an effective tool for selecting scientifically \emph{interesting} sources from a large dataset with minimal human training.  We now apply \protege{} to the entire dataset of \fullcatsize{} extended sources in the MGCLS and look at the sources it considers most \emph{interesting}. 

We applied exactly the same procedure described in \autoref{sec:protege}, using the BYOL features trained on all \fullcatsize{} extended sources\footnote{Note that this includes the \evaluationcatsize{} sources in the evaluation subset, but those human scores were not used.  \protege{} was trained fresh on the full dataset and there is no separate testing or validation phase.}. 

Author ML iteratively trained \protege{} in batches of 10 until 400 sources had been given human scores. \protege{} then gave final scores to the entire dataset.  
After this point, the proportion of sources \protege{} was showing which ML found \emph{interesting} was relatively low. This illuminates an important aspect of how \protege{} training works. In the beginning stages of training, for the first 200 sources, $\sim80\%$ of the objects \protege{} presented were found to be \emph{interesting} (human score of 4 or 5). This is much higher than for the evaluation subset simply because there are more \emph{interesting} sources to find. For the next 100 sources, only 37\% were \emph{interesting}, and that number dropped dramatically to 9\% for the last 100 sources which ML scored. This is an important byproduct of the training process. If the user finds that the fraction of \emph{interesting} sources in each iteration decreases significantly, then further training is likely of marginal value. 

The primary goal of \protege{} is to dramatically reduce the number of sources that must be examined in a very large database to find the ones that are scientifically \emph{interesting}. After training, \protege{}'s performance on this task can be evaluated in the current experiment by comparing images of the top 260 (4.2\%) scorers (\autoref{fig:top_scores}) with 260 randomly selected sources shown in \autoref{sec:random_sources} (\autoref{fig:random_large}) for comparison. The differences are dramatic, showing that \protege{} has been very successful in its primary goal. \href{\fileslink}{A catalog} of coordinates, \protege{} scores and human scores (where available), as well as  \href{\webpagelink}{an online version} of \autoref{fig:top_scores} for all \fullcatsize{} sources can be found at the links provided in the introduction, for follow-up work.

\protege{}'s high-scoring sources show a very diverse set of morphologies, reflecting the diversity of the human training and demonstrating that \protege{} achieved its key design goal. In the rest of this discussion, numbers in parentheses refer to the high scorers in \autoref{fig:top_scores}. Numbers in bold refer to sources highlighted in  \citet{Knowles2022}, as discussed further below.
  
Prominent among the high-scoring sources from \protege{} are spiral galaxies (4, 16, 31, \textbf{47}), head-tail or narrow-angle tailed sources (26, \textbf{44}, 45, \textbf{106}). A variety of radio galaxies exhibiting diffuse and disturbed lobes (0, 2, 6, 8) are also seen, some far from the centers of the target clusters (32, 83, 91). Particularly dramatic are the ``kissing WATs'' featuring apparently two wide-angle or narrow-angle tailed galaxies with extremely bent jets (60). 

\protege{} is also fairly sensitive to diffuse emission, with several high-scoring sources at their cluster centers (1, 10, 19, 22), as well as sources significantly displaced from the centers (11, 21, 25).  These sources, with the exception of source 11, were seen and noted in Table 4 of \citet{Knowles2022}, although they were not among the highlights of the paper.  A newly discovered diffuse source (48), was found close to the center of cluster J0616.8--4748. It has an unusual structure,  with an apparent hole on one side. The utility of \protege{} retrained to specifically detect diffuse emission will be explored in an upcoming paper (Etsebeth et al. in prep).

The enormous diversity among the high-scoring sources is also reflected in  UMAP feature space, shown in \autoref{fig:umap_highlights}, where the high-scoring sources span a large region.  The ability to recover such diverse sources distinguishes \protege{} from other algorithms,  as shown in \autoref{tab:awesome_table}.

\subsection{Case studies}
\label{sec:highlights}
To explore in more detail the selections that \protege{} has made, we present  case studies of two different source types frequently considered to be ``anomalous'' --  X-shaped, and circular-shaped.

\subsubsection{Case study: X-shaped sources}
\label{sec:x_shaped}

\begin{figure*}
\gridline{\fig{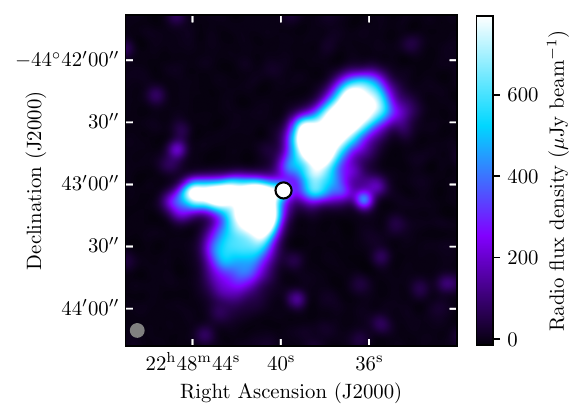}{0.32\textwidth}{(\xclean{})}
          \fig{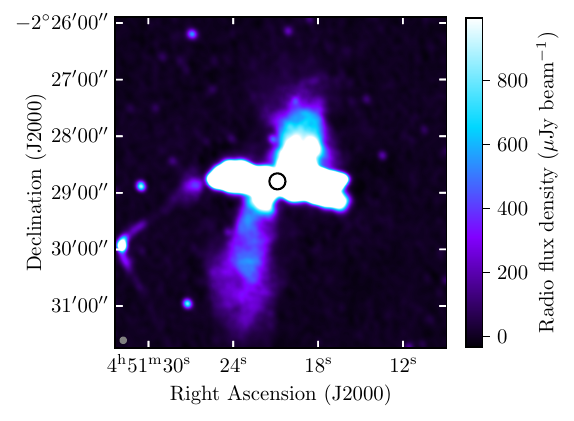}{0.32\textwidth}{(\xdouble{})}
          \fig{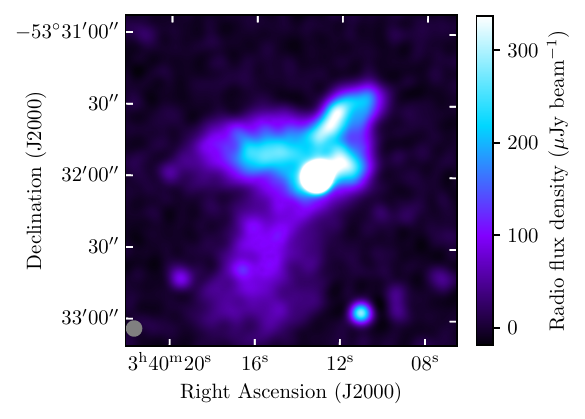}{0.32\textwidth}{(\xlopside{})}
          }

\caption{X-shaped radio galaxies scored highly by \protege{}. The flux density in these images is truncated to 99\% of maximum for the left and right images and 95\% for the central image, to highlight faint features. The potential host, where one is identified, is shown as a white circle and the synthesized beam ($\sim7.6\arcsec$) is shown as a gray circle in the lower left of each image.}
\label{fig:source_x_shaped}
\end{figure*}

X-shaped radio galaxies (XRG) are a fairly rare class of radio galaxies (see \citet{Giri2024} for a recent review). \protege{} was not trained to find this specific morphology, so the sources shown here (\xclean{}, \xdouble{}, \xlopside{}) were identified visually among the high-scorers. 

The most important lesson from this case study is that even among the subset of X-shaped sources, there are significant differences in morphology, and probably the corresponding  physics processes at play. This likely reflects the fact that multiple mechanisms can lead to a general X-shaped source \citep[see e.g.][]{Patra2023}.  

\begin{figure*}
\gridline{
    \fig{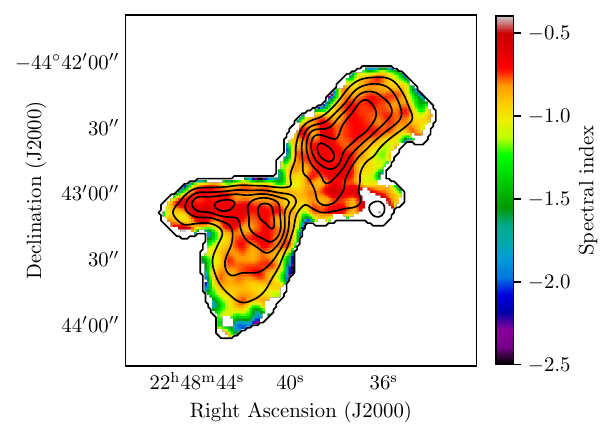}{0.5\textwidth}{(\xclean{})}
    \fig{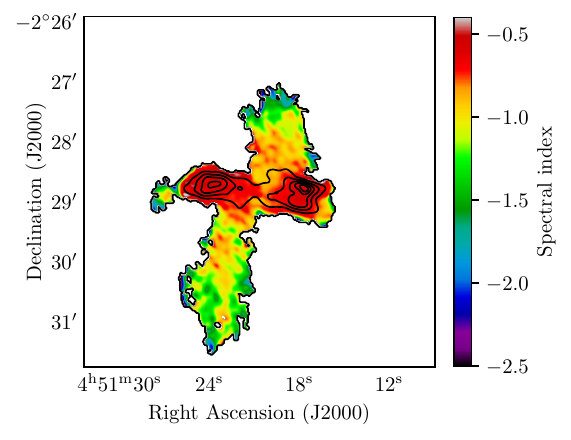}{0.475\textwidth}{(\xdouble{})}
          }

\caption{Spectral index for source \xclean{} (left) and \xdouble{} (right), only including pixels with flux density at least 10 times higher than the rms for the image (4.00 \muJy{} and 7.36\muJy{} respectively) and for which $\sigma_{\alpha}/\alpha < 0.2$. The spectral index is estimated in-band using least squares fitting applied to the twelve sub-band images. Contours are drawn at 5 evenly spaced flux density levels between the minimum and the maximum flux, with background sources excluded.
}
\label{fig:source_x_shaped_alpha}
\end{figure*}

Source \xclean{} (\autoref{fig:source_x_shaped}, left) is a typical XRG, similar to PKS2014-55 in \citet{Cotton2020}, with long straight lobes, indicating the likely orientation of the jets, and short transverse extensions/wings at the back of the lobes, forming an X-shape.\footnote{A possible host for \xclean{}, has a DELVE \citep{DELVE} photo-z of $0.239\pm0.006$, placing it outside the target cluster for this field,  Abell S1063, with a redshift of 0.348.} Using the observed magnetic field structure, \citet{Cotton2020} showed that PKS2014-55 was best understood in terms of hydrodynamic backflows from the ends of the jets, back towards the host galaxy, where they are deflected into an X-shape by the large, oblique hot gas halo of the host galaxy. The structure of \xclean{} appears consistent with this picture. We also find that \xclean{}'s radio  spectra are fairly uniform throughout the source, similar to that of PKS2014-55,  with an average value of $-0.79\pm0.02$ (see \autoref{fig:source_x_shaped_alpha}).

In contrast, source \xdouble{} in the middle panel of \autoref{fig:source_x_shaped}, has a quite different morphology. One set of structures, in the north-south direction, are significantly fainter and longer than the east-west wings. The long, faint structures are not perfectly aligned with the central source and the west wing displays a very unusual double structure in its lobe. The radio spectra are also quite different than those of \xclean{} (\autoref{fig:source_x_shaped_alpha}).  The faint structures steepen away from the host, and on average have much steeper spectra than those of the short wings ($-1.0$, compared to $-0.7$).  These findings are all consistent with a physical model where the jets were originally oriented north-south, and have steepened through synchrotron losses, and then were re-oriented in the east-west direction.  Such reorientations could occur e.g., during a binary black hole merger \citep{Giri2024}.\footnote{Its apparent host has a DES Y3 GOLD \citep{DESGOLD} photo-z of $0.271\pm0.025$, also likely placing it outside the target cluster, PLCK G200.9-28.2 cluster ($z\sim0.220$).}

Source \xlopside{} (\autoref{fig:source_x_shaped}, right panel) has an overall X-shaped morphology, but is unusual in the asymmetry in length and brightness of its opposing structures.\footnote{The host is unclear as the bright source near the center could be a foreground contaminant.} It bears some resemblance to the peculiar XRG of \citet{Kumari2024} which has a one-sided secondary lobe. It is not clear whether either of the two models mentioned above would be sufficient in this case.  It is likely that additional physical processes are involved. The source was too faint to derive spectral maps.

\begin{figure*}
 \includegraphics{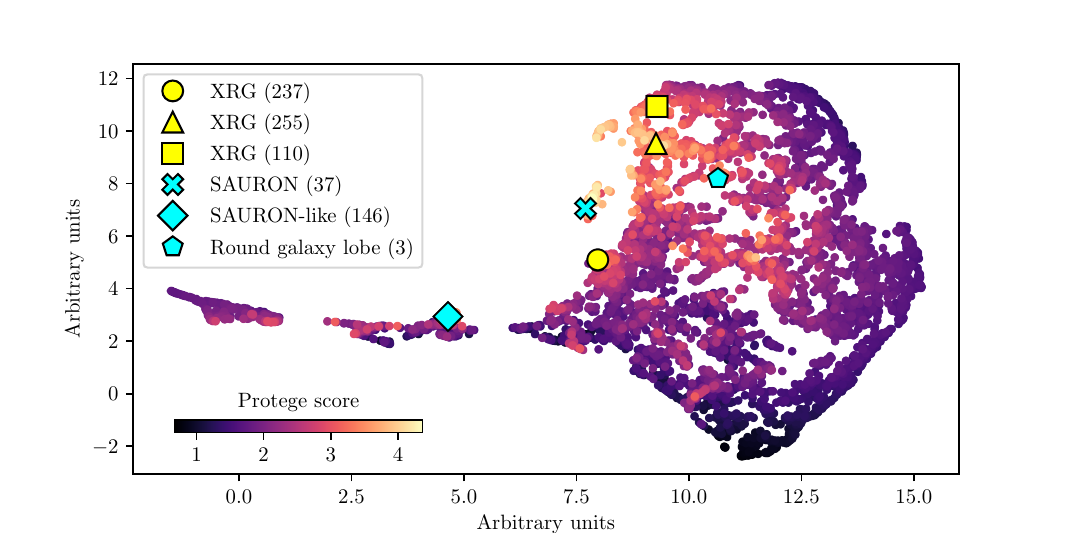}
 \caption{UMAP plot of the BYOL features for the full dataset, colored by the predicted score from \protege{}. Candidate XRGs (source \xclean{}, \xdouble{} and \xlopside{}) and unusually circular sources (\sauron{}, \sauronlike{} and \tail{}) are highlighted. The remarkable diversity even within a single ``class'' of \emph{interesting} source type is reflected by the large distance between them in feature space. Nonetheless, they all lie in regions of interest to \protege{}.}
 \label{fig:umap_highlights}
\end{figure*}

The differences in morphology and spectra between these examples are also reflected in the UMAP feature space (\autoref{fig:umap_highlights}). Thus, if one were interested in finding XRGs, the diversity within this group means that no single similarity search would be sufficient. In addition, since these sources are not isolated or on the edges of the UMAP distribution, they would not be identified through pure anomaly searches in this feature space.   

\subsubsection{Case study: circular shaped sources}

 Circular radio sources, such as supernova remnants, planetary nebulae or face-on spiral galaxies, are well known.  However, a newly discovered class of extragalactic circular radio sources, dubbed Odd Radio Circles \citep[ORCS,][]{Norris2021, Norris2022, Koribalski2021} are not compatible with standard models of jetted radio galaxies. Some ORCs have central galaxies, and so must originate in a spherically symmetric outflow from the AGN \citep{Norris2021}. Other ORCs appear to be associated with diffuse radio lobes from an adjacent AGN  \citep[e.g.,][]{2024PASA...41...24S}. Yet another type of circular object, SAURON, was discovered using \astronomaly{} \citep{Lochner2023} -- a circular structure with quadrilateral brightness asymmetries. Given this new and interesting phenomenon, we looked for circular sources among \protege{}'s high-scorers.
 
\begin{figure*}
\gridline{
    \fig{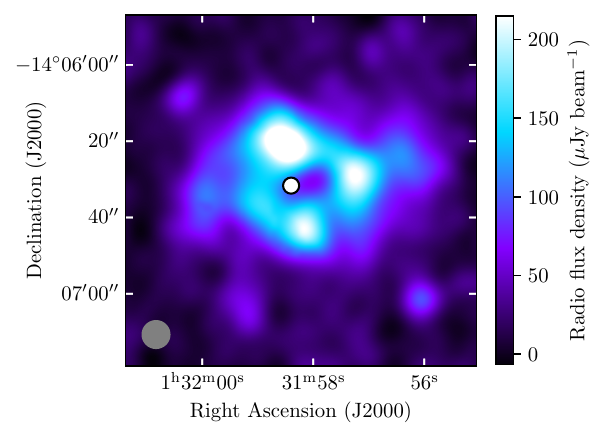}{0.32\textwidth}{(\sauron{})}
    \fig{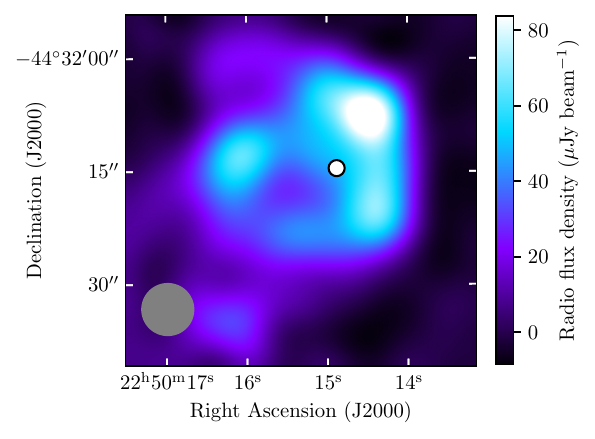}{0.32\textwidth}{(\sauronlike{})}
    \fig{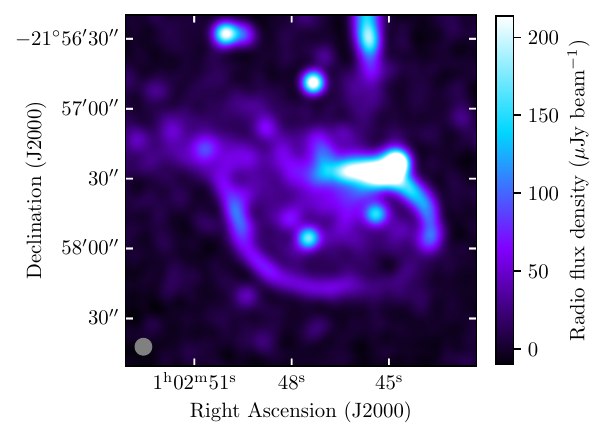}{0.32\textwidth}{(\tail{})}
          }
\caption{Sources detected by \protege{} which are similar to Odd Radio Circles. Source \sauron{} (left) is SAURON, and source \sauronlike{} (middle) is a morphologically similar source discovered by \protege{}. While not a candidate ORC, source \tail{} (right) is an unusually rounded source. The flux density in these images is truncated to 99\% of maximum to highlight faint features and the potential host galaxy (where identified) is indicated with a white circle. The synthesized beam ($\sim7.6\arcsec$) is shown as a gray circle in the lower left of each image.}
\label{fig:source_saurons}
\end{figure*}

 Several high-scoring circular structures were found, including SAURON (\sauron{}, left panel of \autoref{fig:source_saurons}) and another source  (\sauronlike{}, middle panel of \autoref{fig:source_saurons}) with a similar morphology and steep spectrum ($-1.40 \pm 0.07$). It is not clear whether source \sauronlike{} has a host galaxy.\footnote{The optical source in the middle is classified as a star in the DES Y3 GOLD catalog. However, it has no measured parallax in Gaia \citep{Gaia2016, Gaia2023} and DELVE's algorithm instead flags it as a galaxy, with a photo-z of $0.83 \pm 0.04$.} However, if it does, then  source \sauronlike{} has a comparable luminosity to SAURON ($\sim 10^{25}W/Hz$). No other candidate Odd Radio Circles were found in \protege{}'s top 260 recommendations.

\begin{figure}
  \includegraphics[width=\linewidth]{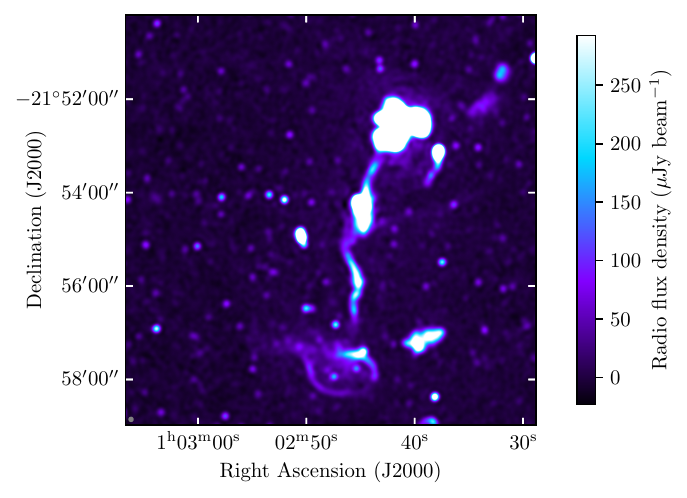}
          
\caption{A zoomed-out image of source \tail{} revealing it as the likely lobe of a large (possibly giant) radio galaxy. The flux density in this images is truncated to 99\% of maximum to highlight faint features and the synthesized beam ($\sim7.6\arcsec$) is shown as a gray circle in the lower left.}
\label{fig:source_tail}
\end{figure}

A final, quite unusual high-scoring circular source (\tail{}) is shown in the right panel of \autoref{fig:source_saurons}. A zoomed-out view  (\autoref{fig:source_tail}) shows that this source may be the southern lobe of a large, highly disturbed radio galaxy. \citet{Randall2010} hypothesized that this could be a background giant radio galaxy at redshift 0.293. Its northern lobe would in that case  coincide with a radio relic at the end of an X-ray plume \citep{2004ApJ...616..157F}, that is located in the foreground target cluster, Abell~133 at z = 0.057. 

These three circular sources are also shown in the UMAP of \autoref{fig:umap_highlights} again illustrating their diversity as they lie far apart in feature space. This seems odd given how visually similar source \sauron{} and \sauronlike{} appear, but visual inspection of a similarity search and these sources' neighbors indicates that the BYOL algorithm has picked up on the diffuse features of source \sauron{} which differ from the fairly smooth, less resolved features of source \sauronlike{}. Again, neither a similarity search nor anomaly selection can replicate the ability of \protege{} to identify this diverse, \emph{interesting} class of sources.

\subsection{Previously identified interesting sources}

Above, we have shown that \protege{} efficiently selected a very diverse set of sources that are of high human interest, as was the goal. We now turn to a related challenge, i.e., how effective is \protege{} in recommending sources that were previously identified as of high interest, completely separate from its training?

For this task, we looked at the \protege{} rank for sources that were highlighted in \citet{Knowles2022}. A number  of the more extraordinary sources discussed there appear in \autoref{fig:top_scores} including a ``phoenix'' (\textbf{7}), the ``double-scythe'' \citep[20, see][]{Chibueze2021} and an unusual source demonstrating ribs and tethers \citep[\textbf{149},][]{Rudnick2021}. A highly morphologically disturbed source at cluster center exhibits remarkable filaments (\textbf{54}, \citet{Rudnick2022}) and we find a new source (195), which could either be another highly disturbed source or even two galaxies, with clear filaments connecting lobes of the source.

In another example, Figure 14 of \citet{Knowles2022} shows a faint radio galaxy exhibiting lateral-edge enhancement (\textbf{135}), a puzzling finding not explained in current radio galaxy models.  \protege{} found another example, ranked extremely high at (5). Equally intriguing are two one-sided sources that appear in the high-scoring sources. One, (133),  has been suggested to result from one of the jets becoming trapped and shocked by the intergalactic medium as two sources merge \citep{Hota2023}. The second source (140) appears to show a normal jet and bright extended lobe, but nothing on the other side (outside of the cutout seen here and by \protege{}). \citet{Kuruhara2023} found that one lobe to have an ultrasteep spectrum, opening the possibility that the other lobe has faded below detectability.

\protege{} successfully gave high scores even to some of the diffuse sources, for which \pybdsf{} is not optimized, including a halo + double relic system (\textbf{29}), ``boomerang'' relics (\textbf{187}) and some dying radio galaxies (\textbf{27, 36}).

To look at this more quantitatively, we checked the 23 examples chosen by \citet{Knowles2022} to highlight in their paper.  These were chosen from hundreds of sources originally suggested for inclusion by team members, of the many thousands of individual sources that were present in the images.  Thus, the highlighted sources represent a set selected to be of the highest interest.

Of these 23, 9 were either too big or too faint to be captured by \pybdsf{} and/or extended beyond the images used as inputs to \protege{}. These included cluster-wide sources such as halos and large peripheral relics. This illustrates a limitation of our current selection. If the objective were to look for faint sources larger than several arcminutes, e.g., a different type of initial source selection than used here would be needed. 

Of the remaining 14, the median \protege{} rank was 70, i.e., among the top 1\% of the sample. 13 of these 14 sources were in the top 2\%, with the remaining source that had been selected by \citet{Knowles2022} for its interesting, but extremely faint diffuse emission. This very high correspondence demonstrates \protege{}'s ability to identify those sources that will be the highest interest to a large team of humans. Author ML, who did the full dataset training here, was not involved in the selection of the sources to highlight in \citet{Knowles2022}, reinforcing this point. 

\newpage
\section{Discussion}
We have demonstrated that \protege{} is a powerful tool for discovering \emph{interesting} sources in the MGCLS dataset. However, there are several avenues for improving \protege{} to ready it for the truly large datasets of the near-future. In this discussion, we highlight some of the strengths and weaknesses of \protege{} with future applications in mind. 

In \autoref{sec:discussion_features} we discuss how the self-supervised deep features derived from BYOL have incredible potential for the field of unsupervised learning, but introduce new challenges that \protege{} is uniquely positioned to address. Ultimately, this self-supervised methodology coupled with small dataset size did not allow for a separate test set. This is unusual for machine learning applications, as discussed in \autoref{sec:discussion_validation}. Preprocessing is another area where improvement is needed, as discussed in \autoref{sec:discussion_preprocessing}, since it led to most of \protege{}'s ``failures.'' Finally, in \autoref{sec:discussion_multi} we speculate how increasingly popular multimodal approaches, combining different data types, could further improve \protege{}.

\label{sec:discussion}
\subsection{Unsupervised learning with deep features}
\label{sec:discussion_features}
As discussed in \autoref{sec:feature_extraction_intro}, feature extraction is generally the limiting factor for machine learning applications, a problem which deep learning has solved for many data types. \citet{Walmsley2022} and \citet{Mohale2024} demonstrated deep-learning derived features can unlock the potential of unsupervised learning for astronomical data, but also revealed a key problem for traditional anomaly detection. UMAP plots in those papers and in \autoref{fig:umap_evaluation} show that while \emph{interesting} sources group together, they rarely fall on the boundary of feature space and the features do not form distinct clusters. This is possibly because the loss functions in deep learning models do not include any specific penalty term for encouraging separation between groups of similar objects. It could also be related to the large number of parameters in deep learning models. 

Traditional anomaly detection methods like isolation forest will fail to detect \emph{interesting} sources that do not fall on the boundary. \protege{} on the other hand, enables the discovery of \emph{interesting} sources even if they are buried deep in feature space. The algorithm only has to find one example to quickly locate any other sources nearby. This is a critical advantage of \protege{} over traditional anomaly detection or the original form of \astronomaly{} which used anomaly detection as a first step (as demonstrated in \autoref{fig:recall}).

However, \protege{} could easily still miss some truly rare sources buried among uninteresting ones and is reliant on the quality of the feature extractor. Therefore, improving the self-supervised methodology, particularly by using foundation models (models trained as general-purpose feature extractors using large datasets), would naturally improve \protege{}'s recommendations.

\subsection{Validation and test sets}
\label{sec:discussion_validation}
It is common in machine learning applications to separate data into subsets for training (to train the algorithm), validation (to help optimize hyperparameters) and testing (to evaluate final performance). This separation prevents overfitting and ensures that models are generalizable. However, our application is not a standard supervised learning problem. We have a two-stage process: feature extraction with self-supervised learning followed directly by an active learning process. This structure, combined with a small dataset, does not easily accommodate the standard training-test data splitting approach.

We ultimately trained BYOL on the full dataset, because more data resulted in better features for the subsequent active learning stage (see \autoref{fig:recall}). At the same time, only the evaluation subset was used to evaluate performance. Although we attempted some hyperparameter optimization and checks for overfitting using validation sets in \autoref{sec:appendix_hyperparams}, the use of the full dataset in our final application means we cannot claim that the trained model would generalize to other radio astronomy datasets. 

However, generalization is not the intention of this work. The goal is to detect as many \emph{interesting} sources as possible in the relatively unexplored and completely unlabeled MGCLS dataset. For use with another data set, the entire procedure, including retraining of the model, would need to be repeated. 

This raises a difference in philosophy between the current experiment and more traditional machine-learning approaches.  Even if we had a large enough dataset to make separate training and test sets, it would not be advantageous to do so. Use of such a hold-out test set would mean ignoring the training set in the search for \emph{interesting} sources, which is our goal. The approach of combining self-supervised and active learning methods is still a new field and understanding the role of validation and test sets in this regime is an important avenue of future research.

\subsection{Preprocessing}
\label{sec:discussion_preprocessing}
\autoref{fig:protege_fails_2} shows a high human-scoring source which \protege{} ranked low. This happens because of the high dynamic range present in these radio images, causing faint emission to be close to invisible. While many transforms have been introduced to mitigate this problem (such as \emph{asinh} used frequently throughout this paper), no universal transform has yet been identified which scales an image perfectly to maintain the necessary detail on multiple flux levels. As found in \citet{Etsebeth_thesis}, machine learning performance is often higher when using a transform which is suboptimal for human identification, leading us to use separate transforms for human and machine tasks. This counter-intuitive result further highlights the importance of understanding the impact of preprocessing in deep learning applications.

In \autoref{sec:appendix_hyperparams}, we found that providing BYOL with linear images without any scaling slightly outperformed \emph{asinh} for this dataset, but that is unlikely to be a general result. An alternative may be to include multiple views of the same image, either with different scaling algorithms applied or in multiple ``slices'' of flux level. As CNNs frequently work with RGB images, the architecture is capable of ingesting this type of data and ensuring that correlations between views of the same image are preserved. This is a useful avenue to explore for any machine learning application for radio or other wavelength data.

\subsection{Adding information to improve \protege{}}
\label{sec:discussion_multi}
Introducing additional information about each cutout could improve \protege{}'s ability to recommend sources the user finds \emph{interesting}. When we further investigated sources of interest identified by \protege{}, we frequently used optical or infrared images. This helped, for example, to distinguish between a spiral galaxy or an unusual source like SAURON. Frequently, the larger context around the source is necessary for understanding its origins (\autoref{fig:source_tail} for example). \autoref{fig:source_x_shaped_alpha} shows how useful spectral information can be, if it is available, in determining the physics of a particular system. In theory, if multiwavelength, spectral or polarization data are available, as well as larger cutouts for each source, these can all be introduced as additional channels for each image before training BYOL. Alternatively, separate feature extractors can be trained and then later combined before passing the features to \protege{} (this is particularly useful if additional data is not available for all sources). In general, multimodal foundation models, deep learning algorithms trained using self-supervised learning on large amounts of different types of data, have proven to be incredibly effective for a number of complex tasks \citep{Paul2024}. This is thus a critical avenue of research for the era of modern astronomy, dominated by multiwavelength surveys.

\newpage
\section{Conclusions}
\label{sec:conclusions}
In this paper, we have made use of a new implementation of machine learning, \protegefull{}, to make discoveries in the MeerKAT Galaxy Cluster Legacy Survey (MGCLS). We explored the concept of how \emph{interesting} a source is, numerically comparing the results of two authors, one considerably more experienced with radio galaxy work than the other. We find good agreement between authors, and with the highly selected \emph{interesting} sources from \citet{Knowles2022}. This indicates that the concept of what is scientifically \emph{interesting}, based on the visual morphology, is a useful one.  

 We find \protegefull{} outperforms other machine learning approaches, as long as the deep-learning based feature extractor (BYOL) has access to sufficient data. The key metric is the recall plot (\autoref{fig:recall}), which asks how much data a human must sift through before detecting a significant number of high human-scoring sources.
 
 With \protege{}, a user only has to search a small amount of data in order to make potentially significant discoveries. We also found examples of sources that \protege{} recommends and humans missed, which highlight that the algorithm has learned something somewhat more fundamental about what makes a source \emph{interesting} to us than the specifics of the human training. 

Many of the sources featured in the original MGCLS paper \citep{Knowles2022} are ranked high by \protege{} but it has discovered many more. We selected a few specific sources to examine in more detail: three candidate X-shaped radio galaxies, SAURON (discovered using a previous version of \astronomaly{} and rediscovered here), a second candidate SAURON and an unusual ring-shaped lobe of a galaxy. These sources, which would be considered \emph{interesting} anomalies by most scientists, demonstrate the diversity of sources recommended by \protege{} and also the richness and discovery potential of modern radio datasets like MGCLS.

Aside from performance, \protege{} has several conceptual advantages over other methods:
\begin{itemize}
    \item Blind methods aimed at finding unusual sources, such as complexity measures and traditional anomaly detection, inevitably include many uninteresting sources such as artefacts. The active learning step introduced in \astronomaly{} mitigates this contamination.
    \item Even with active learning, anomaly detection methods can fail in complex feature spaces such as \autoref{fig:umap_evaluation}. By instead reframing the problem in terms of learning to make recommendations, \protege{} can locate sources of interest buried amongst other sources.
    \item While we focused on developing a general recommender, \protege{} has the flexibility to be tuned to any source of interest including specific classes of source. Although it cannot compete with supervised algorithms if large amounts of training data are available, \protege{} has the advantage of being able to learn from very few examples. 
    \item The idea of a recommender rather than an anomaly detector naturally lends itself to collaborative and citizen science applications, where multiple users' scores could be combined and used as training data to improve \protege{} and provide recommendations to a community of users.
    \item The combination of BYOL-based features and \protege{} can also be applied to image data of any wavelength and wavelengths can even be combined before training BYOL.
\end{itemize}

There are several future research directions that could improve \protege{}. The first and simplest is to increase the size of the training set for the feature extractor, perhaps exploring the use of radio-specific foundation models such as described in \citet{Slijepcevic2023}, fine-tuned onto the dataset of interest.

We also discussed in \autoref{sec:discussion} the need to improve preprocessing, specifically in scaling images with high dynamic range, as well as the rich potential in combining other sources of information, such as multiwavelength data, spectral index maps, polarization and zoomed-out images for context, in improving \protege{}'s recommendations.

A number of sources highly ranked by  \protege{} appear immediately to the authors to be worthy of further science investigation.  We encourage readers to examine \autoref{fig:top_scores} to identify their own sources to pursue, with coordinates and MGCLS target fields \href{\fileslink}{available online}.  We further encourage the use of the  \href{\codelink}{\protege{} software}, to facilitate discoveries in other large and rich astronomical datasets. 

\section*{Acknowledgements}

The authors thank Mike Walmsley and Micah Bowles for useful comments on the draft.
ML acknowledges support from the South African Radio Astronomy Observatory and the National Research Foundation (NRF) towards this research. Opinions expressed and conclusions arrived at, are those of the authors and are not necessarily to be attributed to the NRF.

%Meerkat
MGCLS data products were provided by the South African Radio Astronomy Observatory and the MGCLS team and were derived from observations with the MeerKAT radio telescope. Access to the variety of MGCLS data products is described in \cite{Knowles2022}. The MeerKAT telescope is operated by the South African Radio Astronomy Observatory, which is a facility of the National Research Foundation, an agency of the Department of Science and Innovation.

% Gaia
This work has made use of data from the European Space Agency (ESA) mission
{\it Gaia} (\url{https://www.cosmos.esa.int/gaia}), processed by the {\it Gaia}
Data Processing and Analysis Consortium (DPAC,
\url{https://www.cosmos.esa.int/web/gaia/dpac/consortium}). Funding for the DPAC
has been provided by national institutions, in particular the institutions
participating in the {\it Gaia} Multilateral Agreement.

%ESAsky
This research has made use of ESASky, developed by the ESAC Science Data Centre (ESDC) team and maintained alongside other ESA science mission's archives at ESA's European Space Astronomy Centre (ESAC, Madrid, Spain).

% CDS
This research has made use of the SIMBAD database, CDS, Strasbourg Astronomical Observatory, France. This research has made use of Aladin sky atlas, CDS, Strasbourg Astronomical Observatory, France.

%Data Lab
This research uses services or data provided by the Astro Data Lab, which is part of the Community Science and Data Center (CSDC) Program of NSF NOIRLab. NOIRLab is operated by the Association of Universities for Research in Astronomy (AURA), Inc. under a cooperative agreement with the U.S. National Science Foundation.

%DES
This project used public archival data from the Dark Energy Survey (DES). Funding for the DES Projects has been provided by the U.S. Department of Energy, the U.S. National Science Foundation, the Ministry of Science and Education of Spain, the Science and Technology FacilitiesCouncil of the United Kingdom, the Higher Education Funding Council for England, the National Center for Supercomputing Applications at the University of Illinois at Urbana-Champaign, the Kavli Institute of Cosmological Physics at the University of Chicago, the Center for Cosmology and Astro-Particle Physics at the Ohio State University, the Mitchell Institute for Fundamental Physics and Astronomy at Texas A\&M University, Financiadora de Estudos e Projetos, Funda{\c c}{\~a}o Carlos Chagas Filho de Amparo {\`a} Pesquisa do Estado do Rio de Janeiro, Conselho Nacional de Desenvolvimento Cient{\'i}fico e Tecnol{\'o}gico and the Minist{\'e}rio da Ci{\^e}ncia, Tecnologia e Inova{\c c}{\~a}o, the Deutsche Forschungsgemeinschaft, and the Collaborating Institutions in the Dark Energy Survey.
The Collaborating Institutions are Argonne National Laboratory, the University of California at Santa Cruz, the University of Cambridge, Centro de Investigaciones Energ{\'e}ticas, Medioambientales y Tecnol{\'o}gicas-Madrid, the University of Chicago, University College London, the DES-Brazil Consortium, the University of Edinburgh, the Eidgen{\"o}ssische Technische Hochschule (ETH) Z{\"u}rich,  Fermi National Accelerator Laboratory, the University of Illinois at Urbana-Champaign, the Institut de Ci{\`e}ncies de l'Espai (IEEC/CSIC), the Institut de F{\'i}sica d'Altes Energies, Lawrence Berkeley National Laboratory, the Ludwig-Maximilians Universit{\"a}t M{\"u}nchen and the associated Excellence Cluster Universe, the University of Michigan, the National Optical Astronomy Observatory, the University of Nottingham, The Ohio State University, the OzDES Membership Consortium, the University of Pennsylvania, the University of Portsmouth, SLAC National Accelerator Laboratory, Stanford University, the University of Sussex, and Texas A\&M University.
Based in part on observations at Cerro Tololo Inter-American Observatory, National Optical Astronomy Observatory, which is operated by the Association of Universities for Research in Astronomy (AURA) under a cooperative agreement with the National Science Foundation.


\vspace{5mm}
\facilities{MeerKAT, Blanco, Gaia, Astro Data Lab}


\software{
          Numpy, SciPy \citep{Jones2001}, Matplotlib \citep{Hunter2007}, Seaborn \citep{Waskom2021}, scikit-learn \citep{scikit-learn}, scikit-image \citep{scikit-image},  Pandas \citep{McKinney2010}, Astropy \citep{Astropy1, Price-Whelan_2018}, umap-learn \citep{Sainburg2021}, PyTorch \citep{pytorch} and BYOL-PyTorch \citep{chen2020}}


\appendix

\section{Random sample of sources}
\label{sec:random_sources}
\begin{figure*}
 \includegraphics{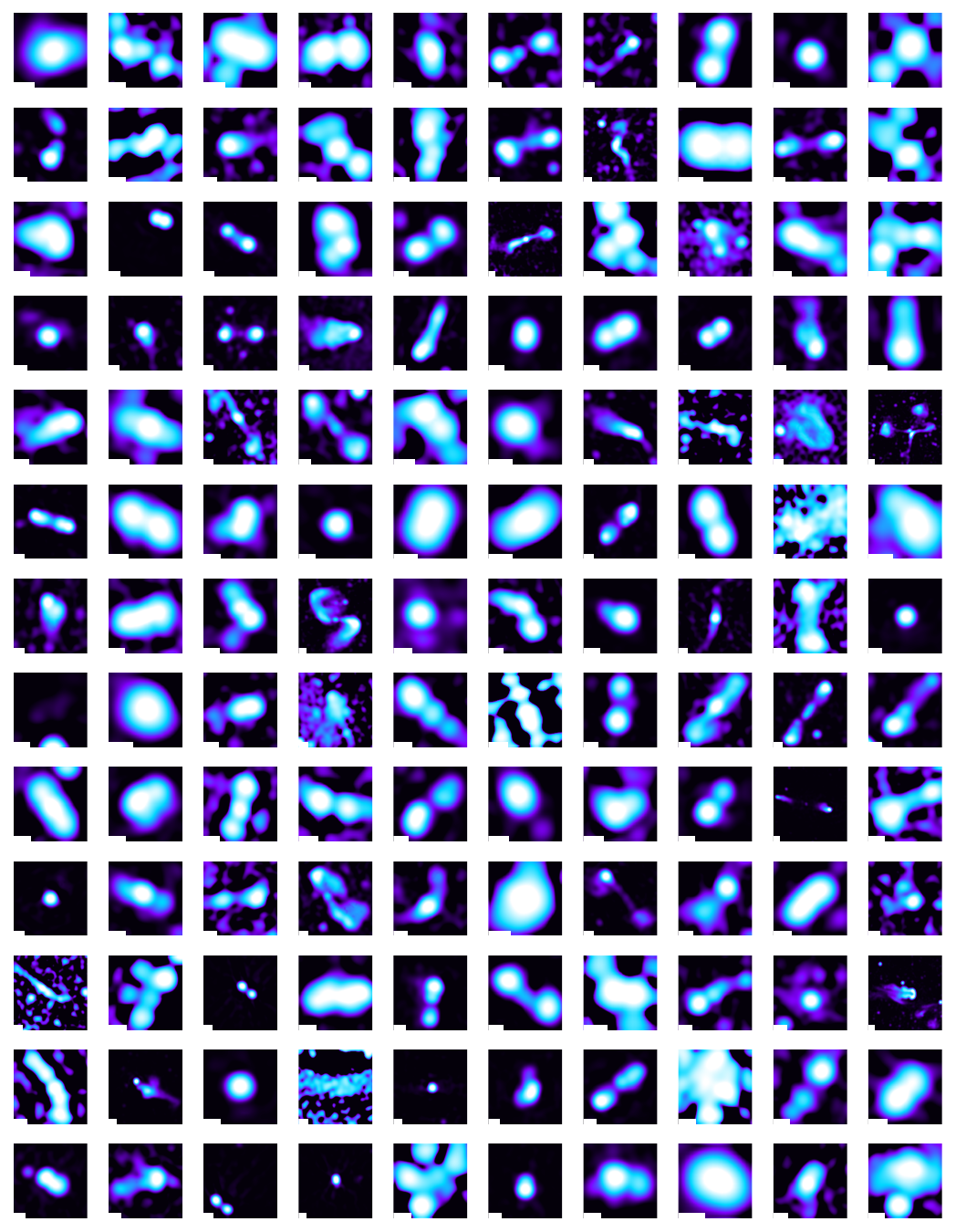}
 \caption{A random sample of sources from the full dataset. Images have been normalized using an $asinh$  scaling with a threshold at 90\% and the beam size ($\sim7.6\arcsec$) is indicated with a bar in the lower left. Figure continued on next page.}
 \label{fig:random_large}
\end{figure*}

The primary purpose of \protege{} was to dramatically reduce the number of sources needing visual inspection to find the scientifically interesting ones. To see how \protege{} performed on this task, we present in \autoref{fig:random_large} cutout images of 260 random sources from the dataset, for direct comparison with \autoref{fig:top_scores}. Note that these random sources are themselves only a small fraction of sources found by \pybdsf{}, viz., those with at least 4 Gaussian components.  The results are dramatic, with \protege{}'s top scorers filled with interesting sources, while the random sources are mostly slightly extended blobs, showing very few features. 

Without \protege{}, a user would have to search through thousands of these sources to make such discoveries as described in \autoref{sec:full_dataset}.

\begin{figure*}
% \ContinuedFloat
 \includegraphics{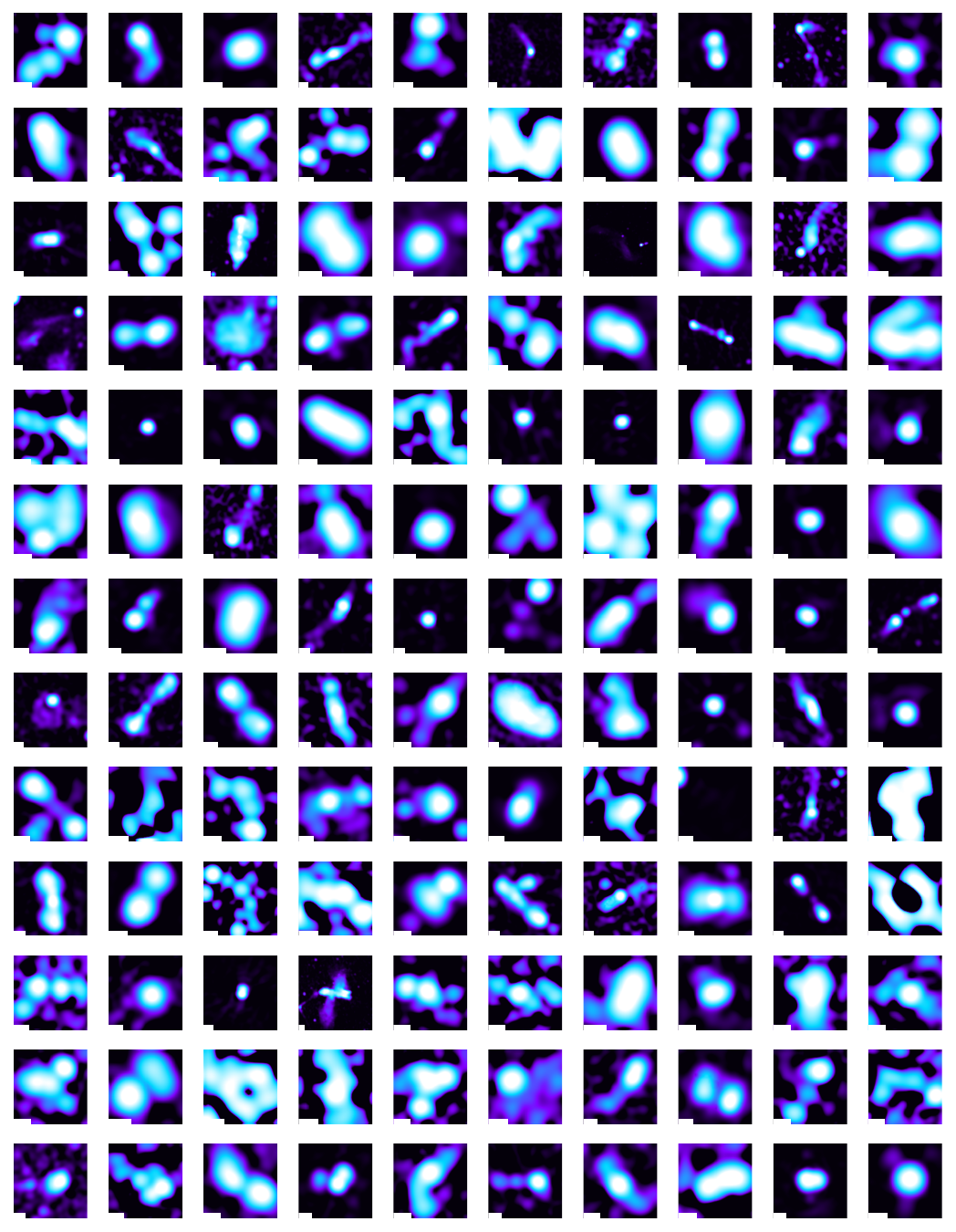}
\end{figure*}

\begin{figure*}
 \includegraphics{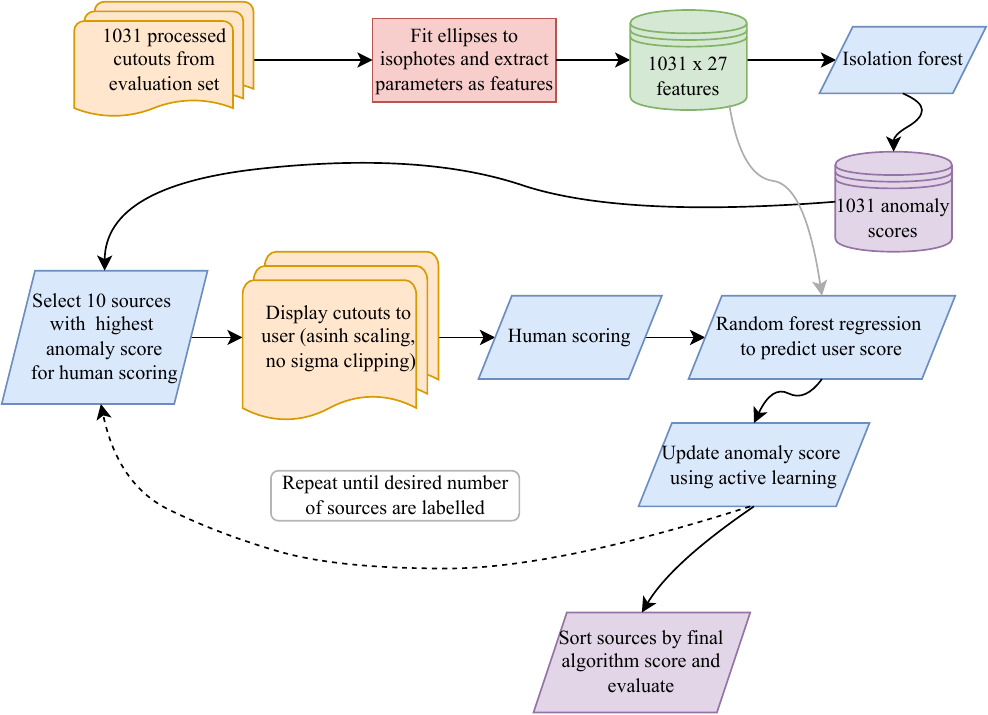}
 \caption{Flowchart describing the original approach taken in \citet{Lochner2021} as applied to the MGCLS data. In contrast to the method described in the main text, features are extracted by fitting ellipses to isophotes (contours of constant brightness) and the standard machine learning algorithm isolation forest is used to indicate anomalies. The active learning process to refine the anomaly scores is a reweighting by combining the isolation forest score, predicted human labels and an uncertainty penalty term. The process is fundamentally different to \protege{} which is trained to focus on regions of interest rather than first filtering through an anomaly detection algorithm.}
 \label{fig:flowchart_ellipses}
\end{figure*}

\section{Comparison with previous techniques}
\label{sec:appendix_ellipses}
In \autoref{sec:evaluation}, we compared \protege{}, using the BYOL features, with the approach introduced in \citet{Lochner2021}, which uses hand-crafted features and more traditional ``true'' anomaly detection techniques. \protege{} significantly outperforms this approach for the MGCLS dataset, but we include it in this appendix for completeness. \autoref{fig:flowchart_ellipses} shows a flowchart of the method.

We used the same preprocessing steps described in \autoref{sec:preprocessing} to extract cutouts of sources. Then, following \citet{Lochner2021}, we use \opencv{} to extract isophotes - contours of equal brightness - corresponding to 90, 80, 70, 60, 50 and 0 percentiles of the brightness distribution of the pixels. We then fit ellipses to these contours, resulting in a set of parameters for each ellipse. The intuition behind using these as features is that more morphologically complicated (and therefore \emph{interesting} to us) sources will have significant differences in parameters such as the rotation and offset from center of each ellipse, while more simple sources will have ellipses with similar parameters. There are 21 features in the final feature set, described in more detail in \citet{Lochner2021}.

We then apply the anomaly detection algorithm isolation forest \citep[iforest,][]{Liu2008}, which is based on the popular random forest algorithm \citep{Breiman2001}. Iforest works by applying random cuts in feature space, using decision trees, in an attempt to isolate individual objects in a dataset. The idea is that more anomalous sources require fewer cuts to isolate. Iforest allows the sorting of the evaluation subset according to how anomalous each source is.

Finally, we apply the novel active learning technique introduced in \citet{Lochner2021}, which we call the ``neighbor score''. This is also an iterative process, where the top 10 sources are displayed to a user to score and then active learning is applied. The process is repeated until 100 sources have been scored. The neighbor score approach first uses a regression algorithm, in this case random forest, to predict the user score for the full dataset. Then the new score is calculated as:

\begin{equation}
\label{eq:active_learning}
    \hat{S} = S\, {\rm tanh} \big(\delta - 1 + {\rm arctanh}(\widetilde{U}) \big),
\end{equation}
where $S$ is the raw anomaly score returned from iforest (normalized from 0 to 5), $\hat{S}$ is the updated active learning anomaly score, $\delta$ is a distance penalty term, that is large for regions where the user score is not well known, and $\widetilde{U}$ is the normalized ``relevance'' score given by user labeling, computed as:
\begin{equation}
\label{eq:userscore}
    \widetilde{U} = \epsilon_1 + \epsilon_2 \left(\frac{U}{U_{\rm max}} \right),
\end{equation}
where $U$ is the predicted user score, $U_{\rm max} = 5$ is the maximum possible user score and $\epsilon_1 = 0.1, \epsilon_2 = 0.85$  are normalization constants.

The distance penalty term, $\delta$, is given by:
\begin{equation}
    \label{eq:distance_penalty}
    \delta = {\rm exp}\left(\alpha \frac{d}{d_0}\right).
\end{equation}
where $d$ is the Euclidean distance, in feature space, of the object in question to its nearest human-labeled neighbor, $d_0$ is the mean distance to a labeled neighbor in the dataset and $\alpha$ is a tuning parameter that essentially dictates the trade-off between the anomaly detection scores and the user scores. We set $\alpha$ to 0.1 for this work, significantly smaller than what was used in \citet{Lochner2021}, as we found it has slightly higher performance. This suggests that the user scores are more informative than the raw anomaly scores.

\autoref{fig:recall} shows that although the ellipse-based features, combined with iforest and the neighbor score active learning approach, perform fairly well, they are outperformed by BYOL and \protege{}.

\section{Hyperparameter optimization}
\label{sec:appendix_hyperparams}
In supervised machine learning, hyperparameter optimization is typically done using a validation set (or cross-validation) and evaluating a clearly defined metric. In this paper, we have introduced a number of algorithms with a dizzying array of hyperparameters. Simultaneously, we have the problems of a small dataset and a fairly ill-defined metric (finding \emph{interesting} sources). Truly optimizing all hyperparameters is extremely difficult due to the high variance inherent in the problem. While we spent considerable effort on hyperparameters following best practices, we focused on understanding the impact of the hyperparameters and selecting values likely to be appropriate for this specific dataset. 

We used the evaluation subset introduced in \autoref{sec:evaluation_subset} to perform several experiments on the effect of various hyperparameters. We will not go into detail in what is already a lengthy paper, but recorded the results of these experiments at \url{\fileslink} for an interested reader to explore. We will describe the broad conclusions drawn from these experiments, and repeat the caution that due to the small dataset size, these are not likely to generalize to other datasets. All experiments were performed on an ordinary desktop computer, making use of an NVIDIA RTX 3060 GPU. Training BYOL on the \evaluationcatsize{} sources of the evaluation subset took around 10 minutes. 

\subsection{A preliminary end-to-end pipeline investigation}
\label{sec:params1}
We initially ran the full pipeline from feature extraction to the detection of high human-scoring sources to examine the impact of several hyperparameter choices. Each experiment was performed 50 times and the mean \recall{} (out of the top 100 sources, how many were scored a 4 or 5 by a human) was recorded, as well as its standard deviation, minimum and maximum values. This metric could vary considerably from experiment to experiment, highlighting the challenge of optimizing these hyperparameters.

First we investigated, for the preprocessing stage,  whether to apply linear scaling or $asinh$ scaling and whether or not to include sigma clipping. \autoref{tab:scaling} shows that using linear scaling with sigma clipping produces slightly better results than the other options.

It is standard practice in deep learning to normalize the images to the same mean and standard deviation of the ImageNet \citep{ImageNet} dataset. We find this very slightly improves performance but it is within the noise. We also investigated the impact of the batch size, finding a batch size of 32 significantly outperforms that of 64 or 128.

\begin{figure}
\centering
 \includegraphics[width=\linewidth]{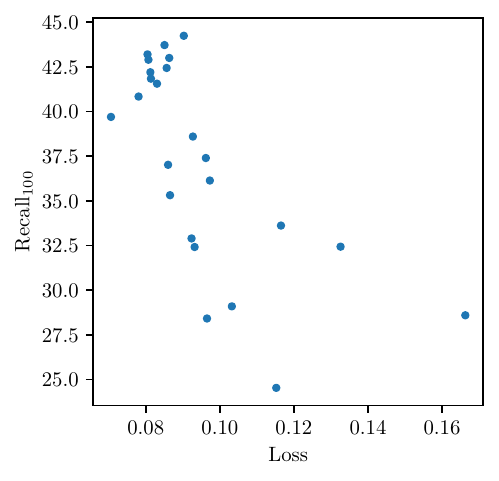}
 \caption{The \recall{} as a function of loss (in BYOL), for a series of experiments with the evaluation subset. There is a very broad correlation between them, suggesting better features results in a better recovery of sources the user finds \emph{interesting}. At the same time, the scatter shows the limitations of our small dataset. }
 \label{fig:loss_recall}
\end{figure}

We explored some amount of variation in the augmentation choices used for BYOL. These were not exhaustive as it was computationally intensive to perform a full ablation study (already done in \citep{Slijepcevic2023}). However in the process, we were able to establish that the \recall{} broadly correlates with the BYOL loss (\autoref{fig:loss_recall}). This implies, somewhat obviously, that better features tend to be more reliable in recovering high human-scoring sources. To separate out the impact of BYOL hyperparameters from those of \protege{}, we then proceeded to explore the impact of augmentations on the loss directly.

\begin{table}
 \centering
 \vspace{20pt}
 \begin{tabular}{cccc}
  \hline
  Scaling & Sigma clipping & \recall{} mean & \recall{} std dev \\
  \hline
  linear & true & 42.88 & 5.30\\
  linear & false & 38.22 & 6.41\\
  $asinh$ & true & 31.00 & 8.26 \\
  $asinh$ & false & 36.14 & 8.03 \\
  \hline
 \end{tabular}
\caption{Effect of preprocessing choices on the ability of \protege{} to recover high human-scoring sources in the evaluation subset. The number of sources given a human score of 4 or 5 in the top 100 sources is referred to as \recall{}. Generally, linear scaling and sigma clipping are slightly preferred.}
\label{tab:scaling}
\end{table}

\begin{table*}
 \centering
 \begin{tabular}{ccc}
  \hline
  Augmentation & Probability & Hyperparameters\\
  \hline
  \texttt{RandomRotation} & $1.0$ & $0\degr-360\degr$ \\
  \texttt{RandomHorizontalFlip} & $0.5$ & N/A\\
  \texttt{RandomVerticalFlip} & $0.5$ & N/A\\
  \texttt{CenterCrop} & 1.0 & 110 pixels \\
  \texttt{RandomResizedCrop} & $1.0$ & $80-100\%$ of image \\
  \texttt{RandomGaussianBlur} & $0.1$ & kernel=15 pixels, $\sigma=10-15$ pixels \\
  \texttt{ColorJiggle} & $0.8$ & All parameters set to $0.5$ \\
  \hline
 \end{tabular}

\caption{Final set of augmentations used for training BYOL, from the \texttt{kornia.augmentation} library, the associated probability of being applied and any relevant hyperparameters. A range is indicated if the hyperparameter is randomly selected from this range at each iteration.}
\label{tab:augmentation}
\end{table*}

\begin{figure*}
 \includegraphics{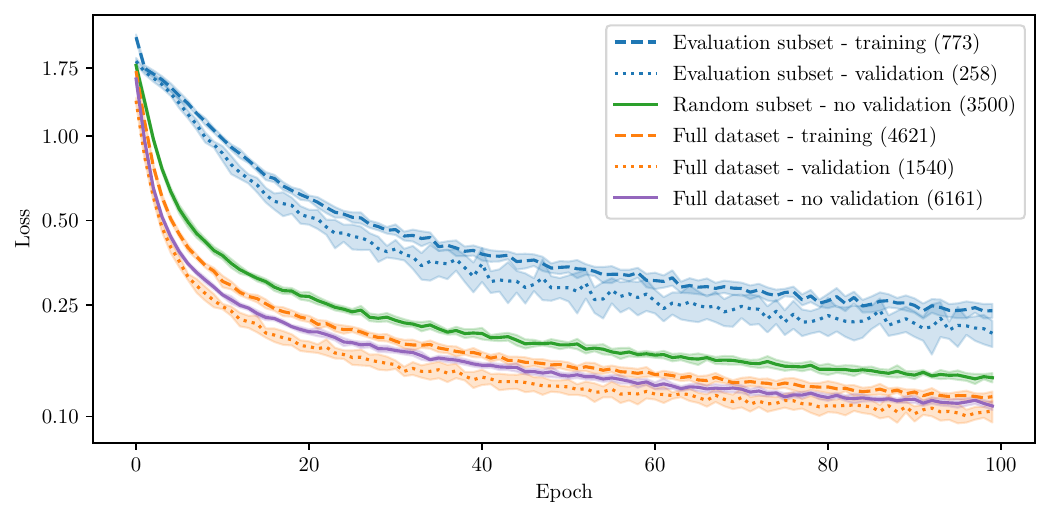}
 \caption{Loss as a function of epoch for various subsets of the data. The solid lines correspond to the subsets used to evaluate performance in \autoref{sec:evaluation} while the dashed and dotted lines split either the evaluation subset or full dataset into training and validation sets to check for overfitting. The number of sources in each subset is given in brackets in the legend. It can be seen that the validation loss shows a similar trend to that of the training, indicating no evidence of significant overfitting. Note that the validation loss is slightly lower than the training loss because dropout is disabled for evaluation. Introducing more data naturally improves the loss function.}
 \label{fig:loss}
\end{figure*}

\subsection{BYOL hyperparameters}
\label{sec:params2}
For each combination of hyperparameters, we trained BYOL ten times, selecting a different random 75\% of the evaluation subset for training each run. We evaluated performance on the validation set loss. Unlike for the experiment in \autoref{sec:params1}, we found that a batch size of 64 performs slightly better than 32. However, since the difference in recall in the first experiment was far more pronounced, we elected to rather use a batch size of 32.

We explored base learning rates from $10^{-5}$ to $10^{-3}$ finding our selected base learning rate of 0.0005 (then scaled by batch size/256) provides a mean validation loss of 0.2 compared to 0.7 for the extreme ends of the search range, justifying our selection of this value.

We used the \textsc{kornia} package \citep{Kornia} to apply augmentations. We did not repeat the augmentation ablation study of \citet{Slijepcevic2023}. However, to ensure the same augmentation choices are applicable for our dataset, we varied some of the individual augmentations and their hyperparameters. We broadly find that for all augmentation hyperparameters explored, the mean validation loss varies by about 25\% with our final choice of augmentations minimizing the loss, with one notable exception. When we removed the augmentation \texttt{ColorJiggle} (the \textsc{kornia} equivalent of Color Jitter), the loss reduced by a factor of three. To test the impact of this choice on the ability of \protege{} to detect high human-scoring sources, we produced two independent feature sets, one with and one without \texttt{ColorJiggle}, using the selected values for all other hyperparameters. \autoref{fig:loss} shows the loss function for various subsets of data, using the final set of augmentations selected. It can be seen that there is no evidence for overfitting (the validation loss does not increase above the training loss) and more training data improves the loss overall. This motivates our choice of using the full dataset to train the BYOL feature extractor when applying \protege{} to the full dataset in \autoref{sec:full_dataset}.

\begin{figure}
 \includegraphics{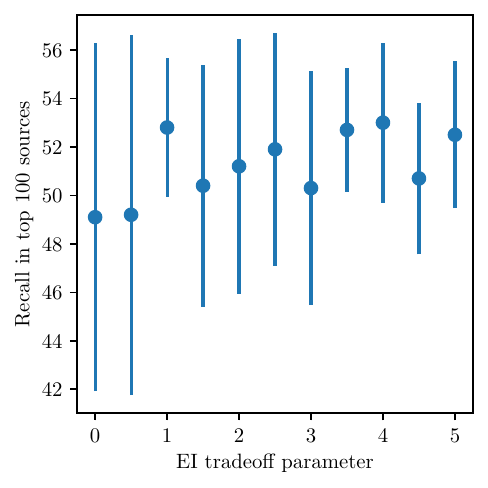}
 \caption{Number of \emph{interesting sources} (human score 4 or 5) detected in the top 100 sources with the highest \protege{} score, after training, as a function of different EI trade-off parameter values. The results are shown for the evaluation subset only. In this optimization exercise, the features are extracted using 75\% of the full dataset for training BYOL with 25\% kept for validation. The error bars are the standard deviation from 10 different training-validation set splits. While the EI trade-off parameter does not significantly affect performance on average, the variance is far lower for values above 0.5.}
 \label{fig:tradeoff}
\end{figure}

\subsection{Optimizing the search for \emph{interesting} sources}
\label{sec:params3}
For this experiment we used two feature sets, switching \texttt{ColorJiggle} on and off, derived using the full dataset and tested only on the evaluation subset. Each experiment is repeated 10 times, with the mean and standard deviation of the \recall{} reported.

We first considered using the active anomaly detection approach of \astronomaly{} introduced in \citet{Lochner2021} with the BYOL features. For a fair comparison with \protege{}, we explored the hyperparameters of isolation forest, local outlier factor, random forest \citep[][used in the neighbor score algorithm of \astronomaly{}]{Breiman2001} and Gaussian processes \citep{Rasmussen2006}. We ended up using the default parameters from the \textsc{scikit-learn} implementations of these algorithms with the following exceptions:
\begin{itemize}
 \item For isolation forest and random forest, we set \texttt{n\_estimators} to 200.
 \item For local outlier factor, we set \texttt{n\_neighbors} to 50.
 \item For Gaussian processes, we use a \texttt{WhiteKernel} and \texttt{Matern} kernel and allow \textsc{scikit-learn} to optimize the hyper parameters.
\end{itemize}

We found that when using the BYOL features, the standard anomaly detection approach does not have high performance. Using the neighbor score active learning method results in a mean \recall{} of 24 when using isolation forest \citep{Liu2008} for anomaly detection and just 14 when using local outlier factor \citep{Breunig2000}. The best performing run from \protege{} in comparison results in a \recall{} of 44. This is why we did not attempt to use standard anomaly detection with these features.

When considering the BYOL features with \protege{} instead, we discovered that retaining the \texttt{ColorJiggle} augmentation slightly improves the \recall{} (39 against 36). We thus retained this augmentation and \autoref{tab:augmentation} summarizes the final set of augmentations used in this work.

We explored different numbers of PCA components finding the impact surprisingly small (\recall{} of 41 against 44). This is why we elected to simply use the variance to dictate the number of PCA components, especially as this will vary depending on which subset of data is used to train BYOL.

Although best practices were followed in hyperparameter optimization and the final set of hyperparameters chosen were the most performant, we conclude that for the majority of hyperparameters, the dataset is too small and the variance too high for them to have significant impact on the results.

\subsection{Effect of the EI trade-off parameter}
\label{sec:ei_tradeoff}
The trade-off parameter $\epsilon$ used in the acquisition function of \protege{} (discussed in \autoref{sec:protege}) could potentially have an impact on performance. In this work, we set this parameter to 3 (halfway between 1 and 5, the minimum and maximum user scores). However, it seemed useful to test for sensitivity to the parameter. We used our final choice of hyperparameters described in \autoref{sec:params3} and trained BYOL 10 times on 75\% of the full dataset, retaining 25\% for validation (as done in \autoref{sec:params2}). We then ran \protege{} with varying values for $\epsilon$, recording the mean and standard deviation in \recall{}. The results, shown in \autoref{fig:tradeoff}, indicate that values of low $\epsilon$ (less exploration) result in more variance. Beyond that, all other values are within the variance and it is difficult to select an optimal value. We thus continued to use an $\epsilon$ of 3 for this work. This parameter could have more impact for larger datasets or different feature sets.

\section{Evaluation of user score regression}
\label{sec:appendix_regression}
\begin{figure}
\includegraphics{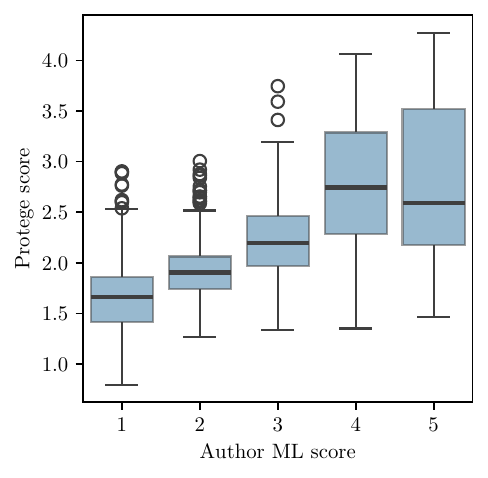}
 \caption{\protege{} predictions vs. author ML's scores for the entire evaluation subset, after training on 100 labels. The data are illustrated using a box and whisker plot, where the median is indicated by a horizontal line in the box, the boundaries of the box are placed at the quartiles of the distribution, the whiskers show the maximum and minimum values barring outliers, given by circles. While the \protege{} scores do not perfectly match the user scores and have high scatter due to a small training set, they nonetheless correlate well enough to quickly locate sources of interest to the user.}
 \label{fig:lochner_vs_protege}
\end{figure}
A key component of \protege{} is its ability to predict the user score for the entire feature space based on a very small amount of training data. We used the evaluation subset (\autoref{sec:evaluation_subset}) to test the performance of the Gaussian process regression algorithm used in this work. \autoref{fig:lochner_vs_protege} compares the user scores provided by author ML with the predictions from \protege{} after training on 100 user-provided scores, in batches of ten per iteration (as described in \autoref{sec:protege}). Despite a large scatter, there is a noticeable correlation between user score and the score from \protege{}.  The scatter is likely due to the very small training sample, e.g., with just 12 examples of sources with a score of five available for the algorithm to learn from. The predictions improve with larger training sets, but even these poor predictions are sufficient to rapidly prioritize sources the user finds \emph{interesting}.

\bibliography{refs}{}
\bibliographystyle{aasjournal}



\end{document}